\documentclass[aps,prd,twocolumn,showpacs,groupedaddress,preprintnumbers,superscriptaddress,amsmath,amssymb,floatfix,longbibliography]{revtex4-1}

\def\nn{\nonumber}

\newcommand{\vect}[1]{\boldsymbol{#1}}

\newcommand{\Avrg}[1]{\left\langle #1 \right\rangle}

\usepackage{cmap}
\usepackage{bbm}

\usepackage{enumerate}
\usepackage{multirow}
\usepackage{float}

\usepackage{xcolor}

\usepackage{color}			
\usepackage{graphicx}
\usepackage[export]{adjustbox}
\DeclareGraphicsExtensions{.png,.jpg}
\graphicspath{}	
\usepackage[section]{placeins} 

\usepackage{physics}
\usepackage{braket}
\usepackage{xparse}	

\usepackage{slashed}
\usepackage{indentfirst} 
\usepackage{enumitem}

\usepackage[colorlinks=true, filecolor=blue, citecolor=blue,urlcolor=blue]{hyperref}
\usepackage{url}
\usepackage{orcidlink}

\date{\today}

\begin{document}

\abovedisplayskip = 4pt
\belowdisplayskip = 4pt
\abovedisplayshortskip = 4pt
\belowdisplayshortskip= 4pt

\title{ Direct-detection constraints on inelastic dark matter with a scalar mediator} 

\author{I.~V.~Voronchikhin \orcidlink{0000-0003-3037-636X}}
\email[\textbf{e-mail}: ]{i.v.voronchikhin@gmail.com}
\affiliation{Institute for Nuclear Research, 117312 Moscow, Russia}
\affiliation{ Tomsk Polytechnic University, 634050 Tomsk, Russia}

\author{D.~V.~Kirpichnikov \orcidlink{0000-0002-7177-077X}}
\email[\textbf{e-mail}: ]{dmbrick@gmail.com}
\affiliation{Institute for Nuclear Research, 117312 Moscow, Russia}

\begin{abstract}
We phenomenologically calculate direct-detection constraints on inelastic dark matter 
(DM) for a scalar portal scenario with leptophilic couplings.
The p-wave velocity suppression of the annihilation cross section of 
scalar-mediated inelastic Dirac DM implies the opening of viable 
regions of DM parameter space in the MeV-GeV mass range.
Xenon-based experiments can provide a constraints on scalar-mediated inelastic fermion dark matter for sub-MeV mass splitting, via endothermic and exothermic spin-independent DM-electron scattering. 
To estimate the relevant constraints, we use public data from the XENON1T, PandaX-4T, and LZ liquid-xenon experiments that measure ionization electron signals.
\end{abstract}

\maketitle

\section{Introduction}

In recent decades, a broad range of cosmological and astrophysical observations has indicated that roughly $85\%$ of the matter content of the Universe cannot be accounted for by the known particle content of the Standard Model~\cite{Planck:2018vyg}. 
The existence of dark matter is inferred from gravitational phenomena on multiple scales, including galactic rotation curves, gravitational lensing, galaxy-cluster dynamics, and the cosmic microwave background~\cite{Bertone:2016nfn}.
By contrast, no non-gravitational interaction between dark matter and SM fields has yet been established experimentally.

The lack of a confirmed non-gravitational signal implies that the particle properties of dark matter remain largely unconstrained beyond its gravitational effects on the visible sector. 
This situation motivates a systematic exploration of dark-matter scenarios capable of accounting for the observed phenomena attributed to dark matter. 
As a result, deriving phenomenological constraints on the parameter space of dark-matter models from current and future experiments helps to narrow the range of viable dark-matter candidates.

Light dark matter in the mass range from $1~\mbox{MeV}$ to $1~\mbox{GeV}$ has attracted considerable attention following a series of sufficiently stringent constraints on heavy dark-matter scenarios~\cite{XENON:2018voc,PandaX-4T:2021bab,LZ:2022lsv}. 
If dark matter in this regime was once in thermal equilibrium with the Standard Model plasma in the early Universe, reproducing the observed relic abundance typically requires an efficient depletion mechanism~\cite{Lee:1977ua,Kolb:1985nn,Krnjaic:2015mbs}.
Such a mechanism is often provided by portal interactions mediated by new fields of different spin.
In such scenarios, the dark sector can communicate with the Standard Model through mediators of spin-0 (e.~g., light hidden Higgs bosons)~\cite{McDonald:1993ex,Burgess:2000yq,Wells:2008xg,Sieber:2023nkq,Guo:2025qes,Voronchikhin:2023qig,Becker:2025vgq,Becker:2026icc,Wang:2025xoq}, spin-1 (e.~g., sub-GeV dark photons)~\cite{Holdom:1985ag,Izaguirre:2015yja,Batell:2014mga,Kachanovich:2021eqa,Lyubovitskij:2022hna,Gorbunov:2022dgw,Claude:2022rho,Wang:2023wrx,Voronchikhin:2024vfu,Gustafson:2025dff,Herrera:2023fpq,Gustafson:2024aom}, and spin-2 (e.~g., massive dark  gravitons)~\cite{Lee:2013bua,Kang:2020huh,Gill:2023kyz,Wang:2019jtk,deGiorgi:2022yha,Voronchikhin:2024ygo,Liu:2026tsl}.

The inelastic dark-matter paradigm was introduced in the context direct-detection experiments where transitions between nearly degenerate states can qualitatively reshape the kinematics of scattering~\cite{Tucker-Smith:2001myb}. 
Originally proposed to explain the anomalous modulation signal reported by the DAMA collaboration~\cite{Bernabei:2013xsa}, inelastic dark-matter paradigm has since developed into a theoretically compelling scenario for sub-GeV thermal dark matter. 

The realization of inelastic dark matter was developed in terms of inelastic fermions, which consist of two states with a mass splitting and dominant off-diagonal interactions~\cite{DeSimone:2010tf}.
The key feature phenomenology arises from off-diagonal interactions between the dark-matter ground state, $\chi_1$ with mass $m_{\chi_1}$, and an excited state, $\chi_2$ with mass $m_{\chi_2} \gtrsim m_{\chi_1}$.
For sufficiently small mass splittings, Boltzmann suppression of the heavier 
state sets in only after freeze-out, and therefore has little impact on the 
total dark-matter relic 
abundance~\cite{Baryakhtar:2020rwy,CarrilloGonzalez:2021lxm}. However, for 
sufficiently large mass splittings, the abundance of the heavier dark-matter 
state begins to be suppressed already by the time of freeze-out~\cite{Izaguirre:2017bqb,Foguel:2024lca}. The relevant parameter space can be excluded by the accelerator based experiments~\cite{Voronchikhin:2025eqm,Gninenko:2026svu,Berlin:2018jbm,Jodlowski:2023ohn,Dienes:2023uve}.

Direct-detection experiments aim to measure rare energy depositions produced by interactions between Galactic-halo dark-matter particles and detector target material~\cite{Goodman:1984dc}. 
In particular, dark-matter particles can scatter off nuclei or electrons in the target material of terrestrial detectors.
The nuclear-recoil channel is the canonical probe for weak-scale dark matter; however, its sensitivity deteriorates rapidly for sub-GeV dark matter, since the recoil energies transferred to nuclei typically fall below experimental threshold~\cite{Schumann:2019eaa}.
Therefore, scattering on atomic electrons provides a well-motivated probe of light dark matter~\cite{Essig:2011nj,Essig:2012yx,Essig:2017kqs,Emken:2019tni}.

In liquid-xenon detectors, this motivates low-threshold analyses based on ionization-only electronic-recoil data, which have become probes of sub-GeV dark matter.
However, in the light dark-matter mass range, the bound-state nature of the initial electron must be taken into account~\cite{Essig:2015cda}.
A realistic treatment of dark-matter-electron scattering requires the shell structure and detector-specific ionization response to be taken into account explicitly~\cite{Caddell:2023zsw}.



In this work, we derive phenomenological direct-detection constraints on the parameter space of thermal, light inelastic fermion dark-matter models.
We considers a leptophilic scalar portal, for which direct scattering off bound electrons provides the dominant channel, whereas the Migdal contribution does not occur at the tree level. We use publicly available electron-ionization data from the XENON1T, PandaX-4T, and LZ experiments. 

For the inelastic scattering process $\chi_i + e^- \to \chi_f + e^-$, we use a following definition of the mass splitting: $\delta = m_f - m_i$.
This definition allows the mass splitting to be either positive or negative.
Additionally, we introduce the relative mass splitting in the following from $\Delta = \delta / m_{\chi_1}$, which is  normalized to  the lighter dark-matter mass.

We note that, the strengthening of the signal in the down-scattering regime, $\chi_2 T \to \chi_1 T$,  is a 
characteristic consequence of inelastic dark-matter kinematics, that can similarly arise for 
mediators of different spin~\cite{Finkbeiner:2009mi}. 
In particular, in the Ref.~\cite{Harigaya:2020ckz} are considered two nearly degenerate 
scalar dark-matter states coupled through a vector mediator. 
The analysis above was specifically focused on the transition from the heavier state to the 
lighter one for a fixed absolute mass splitting chosen to account for the XENON1T electron-
recoil excess~\cite{XENON:2020rca}. 

Next, direct detection of light exothermic iDM was investigated in Ref.~\cite{He:2024hkr} through scattering off atomic electrons and ionization induced by the Migdal effect. 
In this case of pseudo-Dirac dark matter coupled through a vector mediator, $A'$, 
the relative contributions of these channels depend on the dark matter mass 
and mass splitting, as well as on the relative strengths of the dark matter 
couplings to leptons and quarks. Authors of Ref.~\cite{He:2024hkr} map 
their constraints on the thermal target of the iDM 
mediated by hidden vector particle. 
Ref.~\cite{Wang:2025uwh} extended the analysis of exothermic inelastic scalar dark matter coupled through a dark photon to the PandaX-4T data, accounting for both direct electron scattering and ionization induced by the Migdal effect.
However, Refs.~\cite{Wang:2025uwh} derived constraints on the  coupling without focusing on the thermal target.

In the papers~\cite{Harigaya:2020ckz,He:2024hkr,Wang:2025uwh}, the main 
focus is the  signal amplification in the XENON1T and PandaX-4T 
experiments for the downscattering regime, that 
implies the vector mediated iDM.  In the present paper we discuss simplified benchmark scenario with 
leptophilic scalar mediator implying Majorana iDM. 
Since the thermal target  curves differ between vector and scalar mediator 
scenarios, the resulting constraints on the parameter spaces of inelastic dark matter are 
distinct.  In addition, compared with Ref.~\cite{Voronchikhin:2025eqm}, which focused on 
fixed-target searches for $\Delta \simeq \mathcal{O}(0.1)$, the present 
analysis studies the complementary regime $|\Delta| < \mathcal{O}(10^{-5})$ and 
liquid-xenon direct-detection experiments.  
Our analysis  employs modern relativistic ionization  form factors~\cite{Caddell:2023zsw} and the LZ data.

A more systematic description of light pseudo-Dirac dark matter coupled through a vector mediator, including its thermal history, was developed in Ref.~\cite{CarrilloGonzalez:2021lxm}. 
In particular, for a vector mediator, light thermal pseudo-Dirac dark matter with a mass splitting $|\Delta | < \mathcal{O}(10^{-6})$ is constrained by CMB observations~\cite{CarrilloGonzalez:2021lxm}. 
The phenomenology of thermal pseudo-Dirac dark matter with a vector mediator was further developed in Ref.~\cite{Foguel:2024lca} to sufficiently large mass splittings, including interactions associated with baryons and leptons. 
Ref.~\cite{Brahma:2023psr} employed a similar vector-mediator model but focused primarily on the resonant regime, $m_{A'}\simeq 2 m_{\chi}$. 
In the resonant regime, the observed relic abundance can be reproduced with a 
small coupling, thereby avoiding the CMB constraints.

The studies~\cite{CarrilloGonzalez:2021lxm,Brahma:2023psr,Foguel:2024lca}  focused on the parameter space of thermal iDM with a vector mediator.
Our present analysis extends the existing concept of kinematic enhancement of the process  $\chi_2 e \to \chi_1 e$ to the case of a leptophilic scalar mediator.

As we discussed above, the scalar~\cite{Berlin:2018bsc} and vector~\cite{CarrilloGonzalez:2021lxm} portals give rise to different thermal target curves and, as a result, different cosmological constraints. 
In particular, the p-wave 
suppression for the scalar-mediator Majorana iDM does not impose constraints from CMB.
Therefore, for a scalar mediator, the region of parameter space with small mass 
splittings,~$|\Delta| < \mathcal{O}(10^{-6})$, remains relevant for searches in 
direct-detection experiments.
Moreover, the couplings of the scalar mediator  to the heavier lepton generations 
are additionally enhanced compared to its electron coupling.  
Specifically, for $m_{\mu}<m_{\chi_1}$, the model with a scalar mediator and 
Majorana iDM is unconstrained and is a subject for direct-detection experiments   
for mass  splitting~$|\Delta| < \mathcal{O}(10^{-5})$.

This paper is organized as follows. 
In  Sec.~\ref{sec:BenchModels} we  discuss the simplified benchmark model, parameter space for inelastic DM mediated by scalar portal and  direct-detection experiments. 
In Sec.~\ref{sec:SignalInDD} we summarize the general expressions which are used to estimate direct-detection constraints in the cases of exothermic and endothermic reactions.
In Sec.~\ref{sec:ExpectedReach} we discuss the resulted direct-
detection constraints on light scalar-mediated inelastic fermion dark-
matter models. Finally, conclusions are drawn in  Sec.~\ref{sec:Conclusion}.

\section{Benchmark scenarios and experiments \label{sec:BenchModels}}

In this section, we describe the dark-matter models employed in our analysis and the parameter region under consideration.
We also briefly review the main characteristics of the direct-detection experiments considered in this work.

\subsection{Simplified benchmark scenario}

The dimension-5 effective operator that couples a leptophilic scalar dark-sector mediator~$\phi$ to the SM charged-lepton sector  reads as~\cite{Berlin:2018bsc}:
\begin{equation} \label{eq:EFT}
   \mathcal{L}_{\rm eff}^\phi  \supset  \frac{1}{2}(\partial_\mu \phi)^2 -\frac{1}{2} m_\phi^2 \phi^2 - \sum_{\l = e, \mu,\tau}c^{\phi}_{ll}  \overline{l}  l \phi,
\end{equation}
 where we  use the flavor-dependent ratio for the coupling constants~\cite{Chen:2018vkr}:
\begin{equation}
\label{CeeCmumu}
c^{\phi}_{ee}:c^{\phi}_{\mu\mu}:c^{\phi}_{\tau\tau} = m_e : m_\mu : m_\tau.
\end{equation}
The dark matter sector consists of two fermion states~$\chi_1$ and~$\chi_2$, described by the Lagrangian~\cite{Tucker-Smith:2001myb,Batell:2017kty}:
\begin{equation}
\mathcal{L}^{\rm DM}_{\rm kin. term.} \supset  \sum_{i = 1,2} \left[ \frac{1}{2} \overline \chi_i \, i \gamma^\mu  \,\partial_\mu \chi_i
-\frac{1}{2} m_{\chi_i}\, \overline \chi_i \chi_i \right]\, ,
\label{KinTerMajoranaDM}
\end{equation}
where~$m_{\chi_i}$ denotes the physical fermion masses, and we take~$m_{\chi_1}~\lesssim~m_{\chi_2}$ such that~$\chi_1$ is the lightest stable DM candidate.

We focus on the effective benchmark Lagrangian involving a leptophilic scalar dark-sector mediator that couples to a pair of  fermions as~\cite{Dreiner:2008tw,Wang:2025xoq}
\begin{align}\label{eq:EffLagrangianScalarMEDMajoranaiCDM}
\mathcal{L}_{\rm  eff.}^{\rm (+)DM} 
& \supset 
 \text{Re}( \lambda^{\phi}_{\chi_1 \chi_2})  \overline{\chi}_1 \chi_2 \phi.  
\end{align}
 In our simplified scenario,  only off-diagonal terms contribute to  the effective interaction~\cite{Voronchikhin:2025eqm,Krnjaic:2025zjl,DallaValleGarcia:2024zva}, we use 
 the standard notation, $\alpha_{\rm iDM} = (\text{Re}[ \lambda^{\phi}_{\chi_1 \chi_2}])^2/(4\pi)$, for the dark sector fine structure constant.

\subsection{Relic abundance of inelastic dark matter\label{sec:RelAbCalcEq}}

 In this subsction, we discuss the freeze-out mechanism, 
assuming a kinetic and chemical equilibrium between DM and the SM thermal bath in the early Universe~\cite{Griest:1990kh,Edsjo:1997bg}. 
As the Universe expands, dark matter departs from thermal equilibrium and becomes depleted from the thermal bath.
Furthermore, the depletion mechanism via portal interactions leads to observed density of dark matter~\cite{Berlin:2018bsc,Foguel:2024lca,Krnjaic:2025noj}.

The current value of the cold DM relic abundance obtained from 
the Planck 2018 combined analysis is~\cite{Planck:2018vyg,Husdal:2016haj}:
\[
    \Omega_c h^2 = 0.1200 \pm 0.0012.
\] 
The  relic density in the case of the co-annihilation channel~$\chi_{1}\chi_{2}~\to~\ell^{+}\ell^{-}$ is estimated to 
be~\cite{Srednicki:1988ce,Kolb:1990vq}:
\begin{equation}\label{eq:RelDensDM1}
    \Omega_{c} h^2
 \propto  
    \left( \, \, \int\limits_{x_f}^{\infty} \frac{\Avrg{\sigma_{\rm eff} v}}{x^2} dx \, \right)^{-1}, 
\end{equation}
where $\Avrg{\sigma_{\rm eff} v}$ is a effective thermally averaged co-annihilation cross section and $x~=~m_{\chi_1}/T$ is a ratio of DM mass to the temperature of the SM plasma. 
A more detailed discussion and explicit expression of the thermally averaged co-annihilation cross section for the considered benchmarks can be found in the Ref.~\cite{Voronchikhin:2025eqm}.

It should  be emphasized that we focus on relative mass splittings $|\Delta| \ll 1/20$.
In this regime, Boltzmann suppression of the heavier dark-matter state occurs after the chemical decoupling of the visible and dark sectors, at $T \simeq m_{\chi} |\Delta|$~\cite{CarrilloGonzalez:2021lxm}.
Thus, for the relative mass splittings considered here, the thermal relic abundance curves for inelastic dark matter coincide with those of the elastic case.

The additional energy injection of DM annihilation in the early Universe can leave an imprint on the measured anisotropy and polarization spectra of cosmic microwave background~\cite{Slatyer:2009yq}~(CMB).
However, straight-forward constraints from CMB for the considered dark-matter model with a scalar mediator can be weakened due to the p-wave annihilation. 
Thus, inelastic fermion dark matter with a scalar mediator can have an unconstrained  region of parameter space similar to the case with a vector 
mediator~\cite{Berlin:2023qco}.

After the process $\chi_1\chi_2~\to~\ell^+\ell^-$ becomes suppressed, inelastic fermion dark matter chemically decouples from the visible sector at a temperature~$T_f~\simeq~m_{\chi_1}/20$, which fixes the total dark matter abundance. 
However, in the presence of a scalar leptophilic mediator, the processes~$\chi_2\ell~\to~\chi_1\ell$ and $\chi_2\chi_2~\to~\chi_1\chi_1$ can modify the fraction of the heavier dark-matter state.
In particular, the kinetic decoupling temperature can be estimated at the order-of-magnitude level as~$T_{\rm kd}^{\chi}~\simeq~\mathcal{O}(m_e)$, due to the start of electron depletion in the early plasma~\cite{CarrilloGonzalez:2021lxm}.
For small mass splitting, the chemical decoupling between the heavy and light dark-matter states, occurring at temperature $T_{\rm ch}^{\chi}$, is controlled by the process $\chi_2\chi_2~\to~\chi_1\chi_1$~\cite{Foguel:2024lca}. 

It is also worth emphasizing that coannihilation is 
suppressed relative to the process 
$\chi_2 \chi_2 \to \chi_1 \chi_1$ by the coupling 
between the visible and hidden sectors, particularly 
for the benchmark value $\alpha_{\rm iDM} = 0.5$. 
Moreover, one can estimate that the channel 
$\chi_1 \chi_1 \to \chi_2 \chi_2$ is effectively open 
for hidden-sector temperatures satisfying 
$T_{\chi} \gtrsim \mathcal{O}(4|\Delta| m_{\chi_1}/3)$, where we have used $\sigma_{\chi_1 \chi_1 \to \chi_2 \chi_2} \sim \sqrt{v^2 - 8|\Delta|}$ 
and $\langle v^2 \rangle = 6 T_{\chi}/m_{\chi_1}$. 
Nevertheless, this process can also proceed at lower 
temperatures through the high-velocity tail of the 
dark matter distribution. We therefore adopt the 
standard assumption of Ref.~\cite{Baryakhtar:2020rwy} 
that $T_{\rm ch}^{\chi} < T_f$. The fraction of the 
heavier state is given by $f_2 = e^{-|\Delta| m_{\chi_1}/T_{\chi}}/(1+e^{-|\Delta| m_{\chi_1}/T_{\chi}})$, as long as the process $\chi_1 \chi_1 \to \chi_2 \chi_2$ remains efficient~\cite{CarrilloGonzalez:2021lxm}.

For $T \lesssim T_{\rm ch}^{\chi}$, the ratio of number densities of the dark-matter particles in excited and grounded states can be treated as constant for the sufficiently small mass splittings~\cite{Brahma:2023psr}.
In particular, for small mass splittings and a vector mediator, the number densities of the excited and ground states remain comparable at late times~\cite{Foguel:2024lca}.
It should also be noted that, for small mass splitting $|\delta| < 2m_e$, decays of the heavy dark-matter state into visible-sector particles are suppressed~\cite{CarrilloGonzalez:2021lxm,Berlin:2023qco}.
In addition, this parameter region is expected to evade searches for the semi-visible mode in accelerator based experiments, since the decay 
channel~$\chi_2~\to~\chi_1 e^+e^-$ is suppressed. 

One can assume that the Boltzmann suppression at temperature $T_{\delta}~\simeq~\delta~\lesssim~\mathcal{O}(10^{-5})m_{\chi_1}$ occurs after the internal dark-sector chemical decoupling,~$T_{\delta}~\lesssim~T_{\rm ch}^{\chi}$, for dark-matter masses in the range from $1~\mathrm{MeV}$ to $1~\mathrm{GeV}$. 
Under these assumptions, the fraction of the heavy state is~$f_{\chi_2}~\simeq~1/2$. 
Otherwise, Boltzmann suppression can reduce the abundance of the heavier state to a negligibly small level.


A more precise calculation of the relevant temperatures at which the two dark-matter states freeze out and evolve in number density-specifically for the inelastic fermion  DM model with a scalar leptophilic mediator we left for future work. The current analysis uses simplified assumptions.

As an approximate approach, one can treat the dark-matter state fractions 
$f_i$ (the relative abundances of the two states) as free parameters. These 
fractions effectively rescale the constraints derived from direct-detection 
experiments: the electron-scattering cross section $\sigma_e$ inferred from 
the data is scaled down by a factor of $f_i$ to account for the fact that 
only a fraction of the total dark matter participates in the inelastic 
scattering process.

In this work, we focus on the parameter space of light thermal dark matter with scalar mediator defined by:
\begin{equation}\label{eq:paramSpace}
    \frac{m_{\chi_1}}{m_{\phi}} = \frac{1}{3},
\;\;
    |\Delta| \lesssim 10^{-5},
\;\;
    m_{e} \lesssim m_{\chi_1},
\;\;
    \alpha_{\rm iDM} = 0.5.
\end{equation}
For this region we  introduce two scenarios:
\begin{itemize}
\item 
    \textbf{Endothermic (up-scattering) scenario}: 
    The abundance of the heavier dark-matter state is negligibly small, implying~$f_{\chi_1}~\simeq~1$. 
    Consequently, down-scattering in direct-detection experiments is suppressed by the small fraction~$f_{\chi_2}~\simeq~0$.
    As a result, up-scattering process provides the dominant channel for the constraints from direct-detection experiments.
\item 
    \textbf{Exothermic (down-scattering) scenario}:
    The abundances of the heavier and lighter dark-matter states are comparable,~$f_{\chi_2}~\simeq~f_{\chi_1}~\simeq~1/2$. In this case, down-scattering provides the dominant channel for constraints from direct-detection experiments.
\end{itemize}
The results obtained within the proposed scenarios allow us to estimate the  sensitivity of direct-detection experiments to each of the scattering channels. 

\subsection{Experiments\label{sec:ExperimentSetup}}

Direct-detection experiments search for dark matter via rare interactions in a terrestrial target, where such interactions can lead to small electron reconstructed energy. Strong background rejection is employed to isolate candidate recoil signals. 

\textbf{XENON1T}. 
The XENON1T experiment is an underground direct-detection search for dark matter operated from 2015 to the end of 2018 with a integrated time of 258.2 days at the Laboratori Nazionali del Gran Sasso in Italy. 
The core of XENON1T is a dual-phase time projection chamber (TPC) containing of two tonnes of liquid xenon, bounded by a grounded electrode at the top and a cathode at the bottom. 
Energy deposited by charged recoils can generate both prompt scintillation (S1) and ionization electron signal (S2).
To achieve an ultralow-background environment, the active detector is shielded by multiple layers, including an approximately 3600 m water-equivalent rock overburden, an active water Cherenkov muon veto, and an additional 1.2 tonnes of LXe surrounding the TPC.
The dominant background contributions arise from~$\beta$-decays of radioactive impurities, mainly   Pb$^{214}$, in the active detector volume; from neutrino backgrounds, primarily CE$\nu$NS; and from $\beta$-decays originating on the cathode.
In particular, the nearly flat background from $\beta$-decays of impurities is reduced through the use of cleaner materials. 
Also, cathode-related backgrounds are suppressed by applying a selection on the S2 signal width.
We use the data from the Ref.~\cite{XENON:2019gfn} which reports an \text{S2-only} data based on ionization electrons. 
The measured \text{S2-signal} energies for electronic recoils span the range from 0.186~keV (150~PE), with lower-energy events excluded because of poorly controlled backgrounds, up to $\simeq~3$ keV (3000~PE). 
Also, the full detector efficiency is already incorporated into the response matrix.
It should also be emphasized that the publicly available XENON1T S2-only data do not include the analysis-specific cathode background.

\begin{figure}[tbh!]
	\center{\includegraphics[scale=0.5]{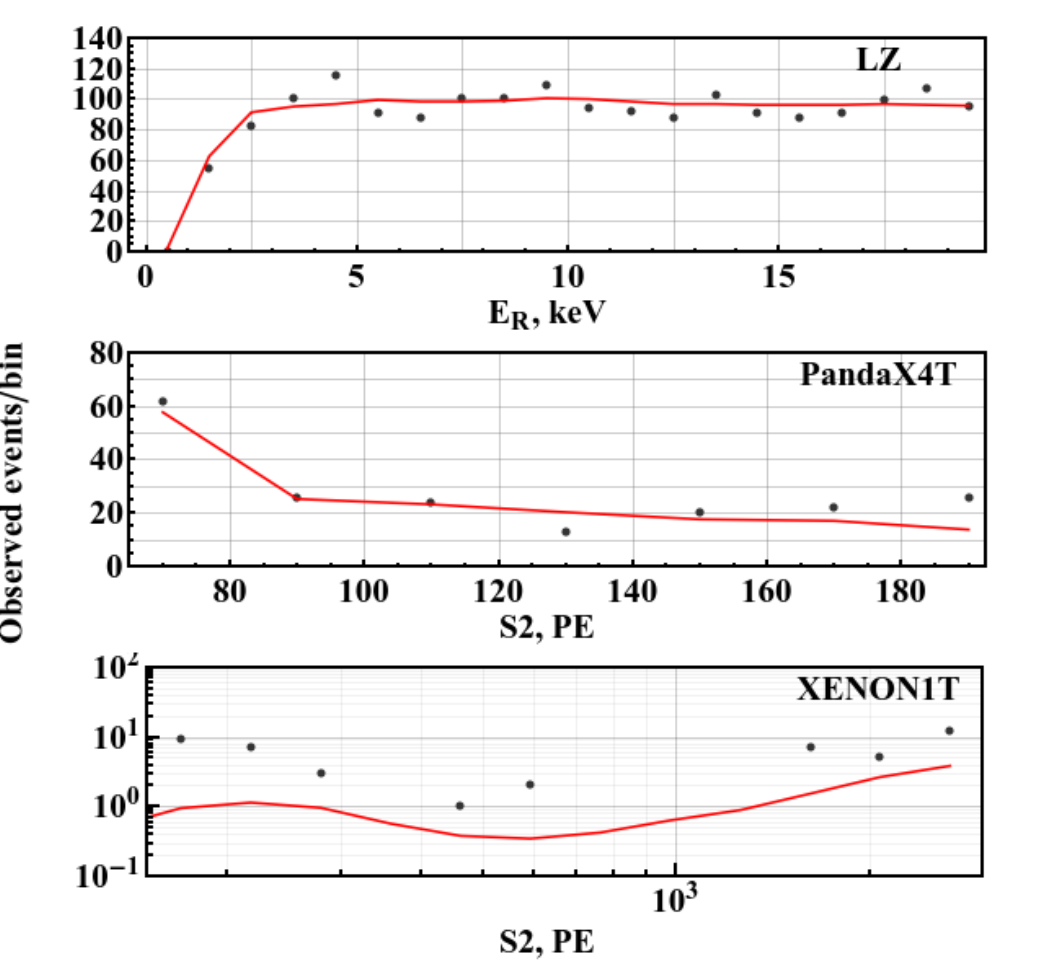}}
\caption{ Event and background public data for XENON1T~\cite{XENON:2019gfn} PandaX-4T~\cite{PandaX:2022xqx}, and LZ~\cite{LZ:2025zpw} experiments. Red line corresponds to case of simulated  background, black points are observed data after imposing all   cuts in experiments.}
	\label{fig:chi1chi2Toll}
\end{figure}

\textbf{PandaX-4T}. 
The PandaX-4T experiment is an underground direct-detection search for dark matter located in B2 hall of the China Jinping Underground Laboratory (CJPL-II) in Sichuan, China. 
Its commissioning run started on November~28,~2020 and ended on April~16,~2021, comprising 95.0~calendar days of stable data taking. 
The detector is a multi-tonne dual-phase xenon TPC with a sensitive target of 3.7~tonnes of LXe contained within a double-vessel cryostat holding 5.6~tonnes of LXe in total. 
The TPC is a cylindrical LXe volume with a cathode at the bottom and gate and anode grids near the surface, providing drift and extraction fields. 
Energy deposited by particle interactions generates
both prompt scintillation photons (S1) in the LXe and a delayed ionization electron signal (S2).
Both signals are collected by PMT arrays at the top and bottom of the TPC.
To achieve a low-background environment, PandaX-4T benefits from an overburden of $\sim$~2.4~km rock (corresponding to $\sim$~6720~m water equivalent) and is surrounded by an ultrapure-water shield in a stainless-steel tank with a diameter of 10~m diameter and a height of 13~m.
The dominant background contribution in the DM-electron ionization-only channel arises from $\beta$-decays of the internal radioactive contaminants. 
We use the public PandaX-4T data based on ionization electrons form Ref.~\cite{PandaX:2022xqx}.
The S2-signal energy range extends from 0.07 keV (60 PE) to 0.23 keV (200 PE) for electronic recoils.  
The lower and upper boundaries are determined by high background rate at very low S2 and by sensitivity to DM search, respectively.

\textbf{LZ}. 
The LUX-ZEPLIN (LZ) experiment is a direct-detection search for dark matter conducted at the Sanford Underground Research Facility (SURF) in Lead, South Dakota, at a depth of 4850 ft (4300 m water equivalent). 
For the 2024 WIMP search, LZ used data collected between 2021 and 2024, corresponding to an exposure 4.2 tonne-years. 
To suppress backgrounds arising from the radioactivity of detector components, the TPC is enclosed within a system consisting of a 2-tonne LXe gamma-tagging detector and an outer detector containing 17.3 tonnes of gadolinium-loaded liquid scintillator, optimized for neutron detection.
These active components are further surrounded by 238 tonnes of ultrapure water, providing additional passive shielding and ensuring a low-background environment. 
Particle interactions in the TPC produce prompt scintillation light (S1) and ionization electrons, which generate secondary electroluminescence (S2). 
The dominant background contribution arises from the $\beta$-decay of Pb$^{214}$, for which a radon-tagging technique is employed.
We use the publicly available LZ data based on ionization electrons from Ref.~\cite{LZ:2025zpw}. The electron-recoil signal energy range extends from 1 keV to 20 keV for electronic recoils.

\section{Signatures of the direct detection \label{sec:SignalInDD}}

In this section, we summarize the general expressions that can be used for 
the estimation of constraints on dark-matter models from direct-detection 
experiments. As mentioned above, we focus on small relative mass 
splittings, which imply $m_{\chi_1} \simeq m_{\chi_2} \equiv m_{\chi}$. We 
consider both exothermic and endothermic processes of dark-matter 
scattering on electrons.

\subsection{Kinematics
\label{subsect:Kinem}}

Let us consider the inelastic scattering of  dark matter off an atomic electron:
$$
\chi_i(p_i) + e^-(p_2) \to \chi_f(p_f) + e^-(p_4),
$$
where the momentum transfer is defined as $q^\mu \equiv p_i^\mu - p_f^\mu$ and the mass splitting is $\delta = m_f - m_i$. 
This process is endothermic ($\delta > 0$) in the case of up-scattering, while down-scattering leads to an exothermic process with $\delta < 0$. 
The momentum transfer depends on the dark-matter velocity $v_{\chi}$ and deposited energy $E_{\rm d}$ as~\cite{Harigaya:2020ckz}:
\begin{equation}\label{eq:limitsTransfMomenta}
	q_{\pm}(v) 
= 
\left|
    	m_{\chi} v_{\chi} 
    \pm 
    	\sqrt{m_{\chi}^2 v_{\chi}^2 - 2 m_{\chi} (E_{\rm d} + \delta)}
\right|,
\end{equation}
which implies the following condition on the dark matter velocity  for exothermic processes with~$|\delta|~<~E_{\rm d}$ and endothermic processes as:
\begin{equation}
v_{\chi}^2~>~2(E_{\rm d} + \delta)/m_{\chi}.
\end{equation}
In particular, for the up-scattering process  one finds the estimate $\Delta~<~(v_{\chi_1})^{2}/2$. 
Assuming that the dark matter population in the Solar System is gravitationally bound to the Milky Way~\cite{Smith:2006ym,Green:2017odb}, the dark-matter velocity is bounded by 
\begin{equation}
v_{\chi}^{\rm max} = v_{\rm esc} + v_{\rm E}~\simeq~2.58~\cdot~10^{-3}.
\end{equation} 
Thus, in the case of up-scattering process, signal events in direct-detection experiments are only possible for a relative mass splitting~$\Delta~\lesssim~\mathcal{O}(10^{-6})$. 

\begin{figure}[tbh!]
	\center{\includegraphics[scale=0.475]{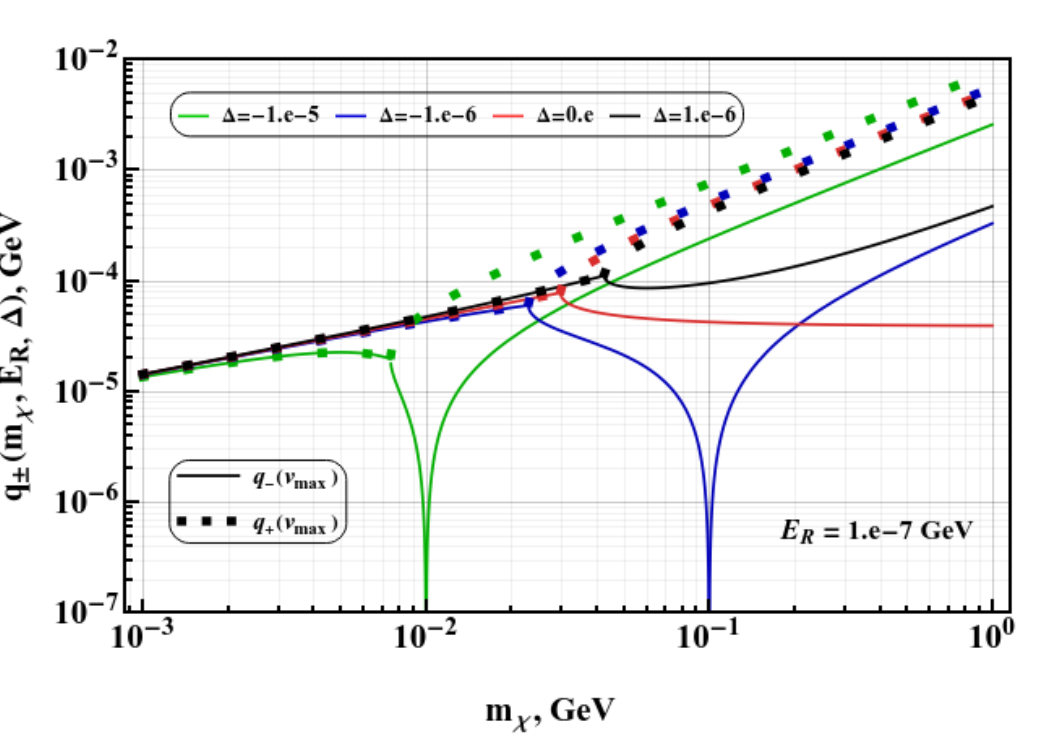}}
\caption{ Transferred momentum limits~\eqref{eq:rangeTransfMomenta} as functions of the DM mass with the fixed deposited energy. Each color corresponds to a different value of the relative mass splitting. The dotted and solid lines denote the upper and lower limits, respectively.}
	\label{fig:qMinMax}
\end{figure}

The minimum dark-matter velocity as function of momentum transfer is:
\begin{equation}
	v_{\rm min}(q)
\simeq
    \left|
    	\frac{E_{\rm d} + \delta}{q}
    +
    	\frac{q}{2m_{\chi}}
    \right|.
    \label{eq:VminDef}
\end{equation} 
Therefore, the lower limit of the dark-matter velocity reaches zero only in the exothermic case with~$|\delta|~>~E_{\rm d}$.
Imposing the dark-matter velocity condition $v_{\rm min}(q) < v_{\chi}^{\rm max}$ leads to the following constraints on momentum transfer:
\begin{equation}\label{eq:rangeTransfMomenta}
   q_{-}(v_{\chi}^{\rm max}) < q < q_{+}(v_{\chi}^{\rm max}).
\end{equation}
The corresponding momentum ranges for different parameter choices are shown in Fig.~\ref{fig:qMinMax}.
It should also be noted that, for the considered processes and the mass splittings satisfying $E_{\rm d} < m_{\chi}|\Delta|$, the limits on the momentum transfer tend to $m_{\chi}\left(v_{\chi}^{\rm max} \pm \sqrt{(v_{\chi}^{\rm max})^2 - 2\Delta}\right)$.
In the region $E_{\rm d} > m_{\chi}(v^{\rm max}_{\chi})^2/2, m_{\chi}|\Delta|$, kinematic suppression occurs, such that $q_{+} \simeq q_{-} \simeq \sqrt{2 m_{\chi} E_{\rm d}}$.
In addition, in the regime,
\begin{equation}
m_{\chi} \simeq E_{\rm d}/|\Delta|,
\label{eq:MassPeak}
\end{equation}
the momentum limits become $0$ and $2 m_{\chi}v_{\chi}^{\rm max}$, which leads to an enhancement in this mass region compared with the elastic case. 
Also, increasing the deposited energy shifts the enhancement region toward larger masses.

\begin{figure}[tbh!]
	\center{\includegraphics[scale=0.475]{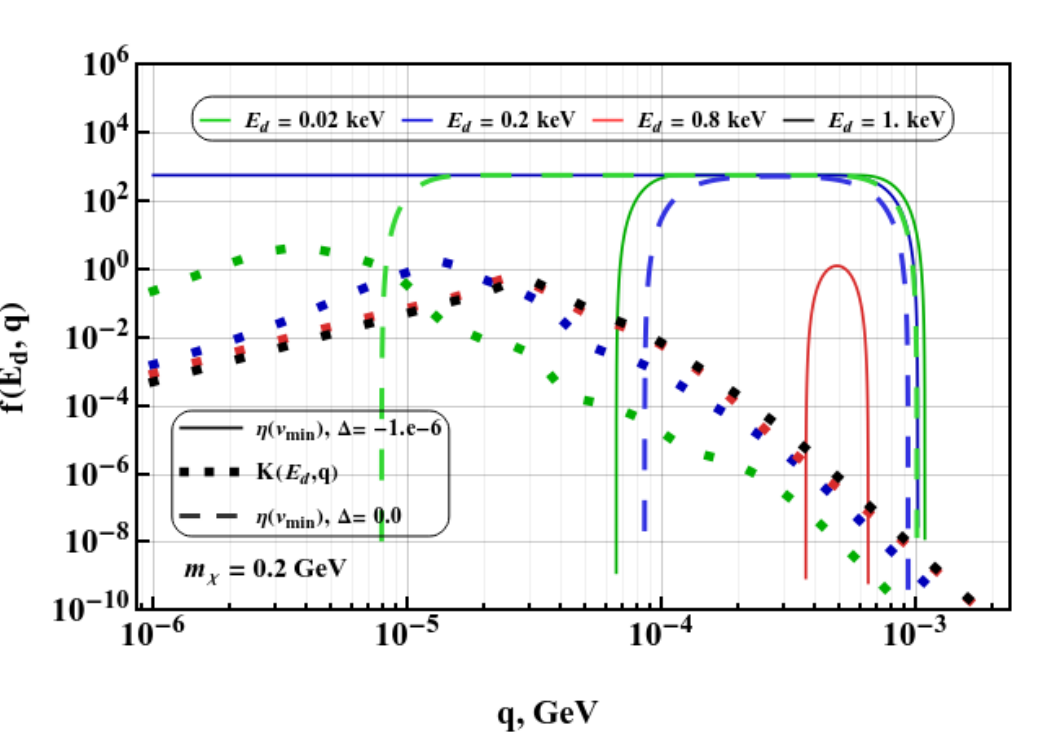}}
\caption{ Total ionization factor and thermally averaged inverse velocity as functions of the transferred momentum for different deposited energies. Different colors correspond to different deposited energies. The quantity $\eta(v_{\rm min})$ is shown for relative mass splittings~$\Delta~=~0.0$ and~$\Delta~=~-10^{-6}$ by dashed and solid lines, respectively. The total ionization factor is shown by the dotted line. }
	\label{fig:IntgrndDiffRate}
\end{figure}

In the light-dark-matter mass regime, the typical deposited energies in 
direct-detection experiments are of order
$$
E_{\rm d} \simeq \mathcal{O}(10^{-2}) - \mathcal{O}(1)~\mbox{keV},
$$
which corresponds to momentum transfers of order $q \lesssim \mathcal{O}(10^{-3}) - \mathcal{O}(1)~\mbox{MeV}$.
Therefore, when light dark matter scatters off a target material, the 
deposited energy can be sufficient to induce inelastic atomic processes, 
and one must account for the bound-state nature of the initial 
electron~\cite{Essig:2011nj}.

The impact of the dark-matter mass splitting on the constraints form direct-detection experiments in the endothermic case becomes important when the 
splitting is comparable to the typical deposited energy,~$|\delta|~\simeq~\mathcal{O}(1)~\mbox{keV}$.
In this case, the up-scattering of inelastic dark matter leads to a smaller 
deposited energy than in the elastic case, resulting in weaker constraints.

\subsection{The experimental reach}

\begin{figure*}[tbh!]
	\center{\includegraphics[scale=1.0]{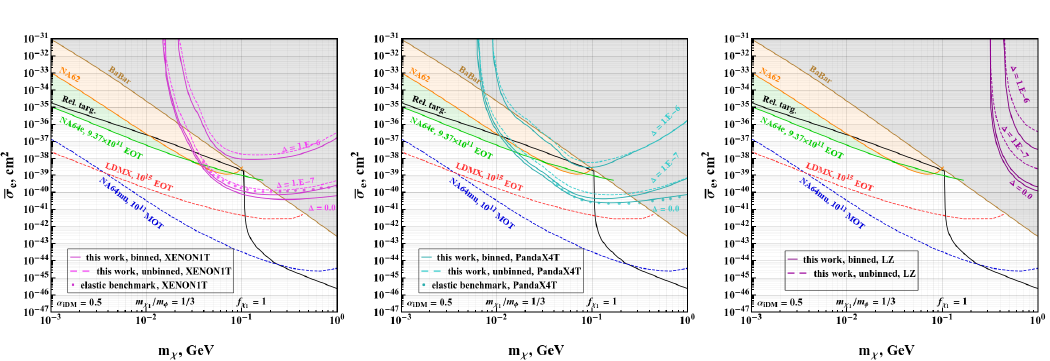}}
\caption{ 
Constraints of effective cross section as function of dark matter mass in cases of benchmark~\eqref{eq:EffLagrangianScalarMEDMajoranaiCDM} and up-scattering~($\Delta~>0$) scenario for different direct-detection experiments. 
Magenta, cyan, purple colors are related by for XENON1T (left panel), 
PandaX-4T (center panel) and LZ (right panel) experiments where solid and 
dashed lines show the binned and unbinned calculations, respectively. 
Magenta and cyan dots indicate the benchmark constraints in elastic cases 
from the Ref.~\cite{XENON:2019gfn} for XENON1T and form the 
Ref.~\cite{PandaX:2022xqx} for PandaX-4T, respectively.
}
	\label{fig:DDconstraintUpScattering}
\end{figure*}

In the general framework of non-relativistic effective field theory~\cite{Krnjaic:2024bdd}, several atomic response functions must be taken into account for scattering off atomic electrons~\cite{Catena:2019gfa,Liang:2024ecw}. 
Let us consider first the effective spin-independent interaction~$(\overline{e} e)(\overline{\chi}\chi)$ that reduces at leading non-relativistic order to the operator~$O_1=\mathbf 1_\chi \mathbf 1_e$. 
 The matrix element for dark-matter scattering off a bound electron admits 
 a simple factorization~\cite{Catena:2019gfa,Liang:2024ecw}:
$$
\mathcal M^{\rm bound}_{i\to f}(\mathbf q)
=
\mathcal M^{\rm free}_{\chi e}(q)
\langle f|e^{i\mathbf q\cdot \hat{\mathbf r}}|i\rangle,
$$
where $\ket{i}$ and $\ket{f}$ denote the initial and final electron states, respectively, and $M^{\rm free}_{\chi e}(q)$ is the matrix element for dark-matter scattering off a free electron. The transition amplitude $\langle f|e^{i\mathbf q\cdot \hat{\mathbf r}}|i\rangle$ encodes the structure of the target material, and its explicit form depends on the normalization convention adopted.
Moreover, at momentum transfers of order $|\vect{q}|^2~\simeq~(\alpha m_e)^2$, the bound-state nature of the initial electron becomes important.
Following the standard approach, we introduce a reference cross section $\overline{\sigma}_e$ and the dark-matter form factor $F_{\rm DM}^2(q)$ as follows~\cite{Essig:2011nj}: 
$$
	\overline{\sigma}_e 
\equiv 
	\frac{ \mu_{\chi e}^2 }
	     { 16\pi m_\chi^2 m_e^2}
	\left|M^{\rm free}_{\chi e}(q)\right|^2_{|\vect{q}|^2=(\alpha m_e)^2},
$$
$$
	F_{\rm DM}^2(q)
\equiv
	\frac{\left|M^{\rm free}_{\chi e}(q)\right|^2}{\left|M^{\rm free}_{\chi e}(q)\right|^2_{|\vect{q}|^2=(\alpha m_e)^2}}.
$$
This representation factorizes the entire transferred-momentum dependence 
of the free-electron matrix element into the dark-matter form factor.

\begin{figure*}[tbh!]
	\center{\includegraphics[scale=1.0]{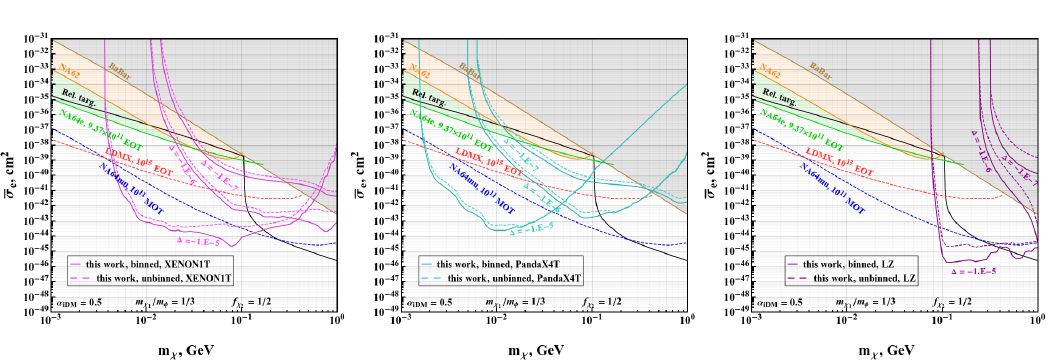}}
\caption{ 
The same as Fig.~\ref{fig:DDconstraintUpScattering}, but for down-scattering~($\Delta~<~0$) scenario.
}
	\label{fig:DDconstraintDownScattering}
\end{figure*}

The differential event rate for dark matter with fraction~$f_i$  is given by~\cite{Caddell:2023zsw}:
\begin{equation}
	\frac{dR}{dE_{\rm d}}
=
	\frac{N_T \rho_\chi f_{i} }{m_T m_\chi}
	\frac{d\langle \sigma v \rangle}{dE_{\rm d}},    
    \label{eq:dRdE}
\end{equation}

where $N_{\rm T}$ is the number of target atoms of mass $m_{\rm T}$, $E_{\rm d}$ is the energy deposited in the target, $\vect{v}$ is the dark-matter velocity in the Earth frame, and $\rho_{\rm DM} \simeq 0.4~\mbox{GeV}/\mbox{cm}^3$ is the local dark-matter density near the Earth~\cite{Read:2014qva,Green:2017odb}.
The thermally averaged differential cross section for dark-matter scattering off a bound electron is~\cite{Essig:2011nj,Caddell:2023zsw}:
\begin{align}
	\frac{d\langle \sigma v \rangle}{d E_{\rm d}}
=  \nn
	\frac{\bar\sigma_e}{2\mu_{\chi e}^2} 
	\int\limits_{q_{-}(v_{\chi_1}^{\rm max})}^{q_{+}(v_{\chi_1}^{\rm max})} 
&
	q 
	\frac{K(E_{\rm d}, q)}{E_{\rm H}}
\\ \cdot &
    \eta\left(v_{\rm min}\right)
	|F_{\rm DM}(q)|^2
	dq,
\end{align}
where $E_{\rm H} = m_e \alpha^2$ is the Hartree energy, $K(E_{\rm d}, q)$ 
is the corresponding total atomic ionization factor, and $\eta(v_{\rm min})
= \left\langle \frac{1}{v} \theta(v - v_{\rm min}) \right\rangle$ is the 
averaged inverse velocity of dark matter over the distribution function 
$f_{\rm TMB}(\vec{v} + \vec{v}_{\rm E})$. In the case of the Standard Halo 
Model, one can use the local truncated Maxwell-Boltzmann distribution 
as~\cite{Savage:2006qr}:
\begin{equation*} 
    f_{\rm TMB}(\vect{v} + \vect{v}_{\rm E})
=
\begin{cases}
    \frac{1}
         { N_{\rm esc}}
    e^{\left(-|\vect{v} + \vect{v}_{\rm E}|^{2}/v_{0}^{2}\right)},
    & |\vect{v} + \vect{v}_{\rm E}| < v_{\rm esc}\\
    0, & |\vect{v} + \vect{v}_{\rm E}| > v_{\rm esc},
\end{cases}
\end{equation*}
$$
    N_{\mathrm{esc}}
=
    \pi^{3/2}v_{0}^{3}
    \left(
        \operatorname{erf}(v_{\rm esc}/v_{0})
    -   \frac{2}{\sqrt{\pi}} 
        \frac{v_{\rm esc}}{v_{0}} 
        e^{-v_{\rm esc}^2/v_{0}^2}
    \right).
$$
where $v_{0}~\simeq~220~\mbox{km}/\mbox{s}$ is the characteristic velocity, 
$v_{\rm esc}~\simeq~544~\mbox{km}/\mbox{s}$ is the escape velocity and 
$\vect{v}_{\rm E}$ is the velocity of the Earth in the Galaxy with 
$v_{\rm E}~\simeq~232~\mbox{km}/\mbox{s}$.
An explicit expression for $\eta(v_{\rm min})$ is provided in the Ref.~\cite{McCabe:2010zh} that is nonzero within the momentum-transfer range defined in~\eqref{eq:rangeTransfMomenta}.
Tabulated values of the total atomic ionization factor for different target materials can be 
found in Ref.~\cite{Caddell:2023zsw}, where this quantity is computed using a relativistic 
Hartree-Fock method. 
We explicitly take into account the momentum-transfer limits in 
order to improve the robustness of the numerical calculations.

Since the relevant particle masses are larger than~$\mathcal{O}(1)~\mbox{MeV}$, the momentum transfer is much smaller than the particle masses. 
Therefore, we work in the regime $t \ll m_e^2, m_\phi^2, m_{\chi_1}^2$. 
After summing over final and averaging over initial internal degrees of freedom, the squared matrix element becomes:
\begin{align}
	\overline{|\mathcal M|^2}
& =
	\frac{ 4 \pi \alpha_{\rm DM} (c_{ee}^{\phi})^2 }{ (t - m_\phi^2)^2 }
	(4m_e^2 - t)
	\left(  (m_{\chi_1} + m_{\chi_2})^2 - t  \right)
\nn \\ & \simeq
	\frac{4 \pi \alpha_{\rm DM} (c_{ee}^{\phi})^2}{(t-m_\phi^2)^2}
	(4m_e^2)
	(m_{\chi_1}+m_{\chi_2})^2
\end{align}
The dark-matter form factor and the effective cross section  read, respectively
\begin{equation}\label{eq:ExpliсMajoranaDMFF}
	F_{\rm DM}^2(t)
=
	\frac{((\alpha m_e)^2 + m_\phi^2)^2}{(|\vect{q}|^2 + m_\phi^2)^2},
\end{equation}
\begin{equation}\label{eq:ExpliсMajoranaDMCS}
	\bar\sigma_e
= 
	\frac{4 \pi \alpha_{\rm DM} (c_{ee}^{\phi})^2}{\pi}
	\frac{\mu_{e\chi_1}^2}
	     {(m_\phi^2 + (\alpha m_e)^2)^2}
	.
\end{equation}
In the parameter region of interest, $m_{\chi}/m_{\phi}~=~1/3$ and $m_e~<~m_{\chi}$, the dark-matter form factor can be set to $F_{\rm DM}^2(t)~\simeq~1$.
However, in deriving the constraints, we explicitly included the form 
factor~(\ref{eq:ExpliсMajoranaDMFF}) in the calculation, and we also 
checked that $F_{\rm DM}^2(t) \simeq 1$ is a reasonable approximation.

The Migdal effect~\cite{Yi:2026fmf} induced by DM-nucleus scattering can provide an additional contribution to 
low-energy electron signals~\cite{Ibe:2017yqa,He:2024hkr}. 
However, in the present work we focus on a leptophilic scalar mediator, for which the 
dominant interaction is the tree-level coupling to electrons. 
For this reason, we restrict our direct-detection analysis to the electron-scattering channel 
and leave a dedicated study of Migdal contributions for future work, specifically for 
hadron-specific mediator scenario.


In our analysis of direct-detection data, we constrain the signal strength using a one-dimensional profile profile-likelihood procedure. The expected number of events in each bin is given by
$$
\mu_i(\bar\sigma_e) = n_{{\rm Bkg},i} +  \bar\sigma_e/(1~{\rm cm}^2) \cdot  n_{{\rm thr},i}(\bar\sigma_e = 1~{\rm cm}^2).
$$
For the full set of bins, we construct the Poisson log-likelihood as
$$
	\log L(\bar\sigma_e)
\! = \!
	\sum_i
	\left[
		    n_{{\rm Obs},i} \ln(\mu_i(\bar\sigma_e))
	\! -\!  \mu_i(\bar\sigma_e)		
    \! - \! \ln\big(n_{{\rm Obs},i}!\big)
	\right].
$$
We obtain the maximum-likelihood estimate of the upper limit by numerically maximization. The profile-likelihood-ratio test statistic is~\cite{Baxter:2021pqo}:
\begin{equation}\label{eq:binnedStatistics}
	q(\bar\sigma_e) = -2\left[\log L(\bar{\sigma}_e)-\log L(\hat{ \bar{\sigma}}_e)\right].
\end{equation}
The upper limit on the parameter is obtained by solving $q(\bar\sigma_e)=1.642$, which corresponds to a one-sided 90\% confidence level in the asymptotic Wilks approximation for a single parameter.

One can calculate number of theoretical signal events by the expression:
\begin{equation}
    N_{\rm sign.} 
= 
    \epsilon \cdot 
    \sum_{j}
    \sum_{i} 
    R_{ij}
    \frac{N_T \rho_\chi}{m_T m_\chi}
	\frac{d\langle \sigma v \rangle}{dE_{\rm d}}({E_{\rm d}}_i)
    d{E_{\rm d}}_i,
\end{equation}
where $\sum_{j}$ and $\sum_{i}$ are sum over event bins and energy discretizations, respectively, $R_{ij}$ is response  matrix. 

For XENON1T~\cite{XENON:2019gfn} and 
PandaX-4T~\cite{PandaX:2022xqx}, we employ the published 
response matrix, which already incorporates the full 
detector efficiency. LZ~\cite{LZ:2025zpw} publicly provides 
the total energy-dependent efficiency and the reconstructed-
energy distributions, but without a response matrix that 
accounts for transitions between different energy bins. We 
therefore adopt the diagonal-response approximation $R_{ij}
^{\rm LZ} = \varepsilon_{\rm LZ} \delta_{ij}$. Accordingly, 
the LZ result should be understood as a public-data recast 
within this approximation.

For an order-of-magnitude estimate of the constraints from direct-detection 
experiments, one can use the  Bayesian  approach and derive limits based 
on~\cite{Magill:2018jla,Magill:2018tbb,ParticleDataGroup:2024cfk}:
\begin{equation}\label{eq:unBinnedStatistics}
    \Gamma(n_{\rm obs}+1, s_{\rm upper} + N_{\rm bkg}) 
= 
    \alpha \Gamma(n_{\rm obs.}+1, N_{\rm bkg}).
\end{equation}
This unbinned treatment, used to derive the signal bound, yields a conservative estimate of the
cross-section limit (shown by dashed lines in Figs.~\ref{fig:DDconstraintDownScattering} 
and~\ref{fig:DDconstraintUpScattering}) and therefore provides weaker constraints.

We use the background-only best-fit spectrum, $n_{\mathrm{Bkg},i}$, published by each collaboration as the background model in our analysis. PandaX-4T and LZ report relative total background uncertainties of about $\mathcal{O}(10)\%$~\cite{PandaX:2022xqx} and $\mathcal{O}(3)\%$~\cite{LZ:2025zpw}, respectively. For XENON1T, a comparable fractional uncertainty is difficult to estimate from the public data owing to the analysis-specific cathode background~\cite{XENON:2019gfn}.

In general, background uncertainties broaden the range of signal strengths compatible with the data and therefore weaken the expected constraints~\cite{Cowan:2010js}. For instance, using a background-only data set at $m_{\rm DM}=4.5\;\mathrm{GeV}$, the DarkSide-50 collaboration demonstrated that uncertainties in the theoretical background spectra and normalizations weakened the expected limit by 15\%–31\%~\cite{DarkSide-50:2023fcw}. By analogy, we expect background uncertainties to affect the limits for the experiments considered here at a similar order of magnitude.

For $\Delta=0$, our XENON1T limits~\cite{XENON:2019gfn} agree with the corresponding collaboration results within a factor of $\mathcal{O}(1)$, while the elastic bounds for PandaX-4T agree with those of~\cite{PandaX:2022xqx} to within $\mathcal{O}(10\%)$. A full treatment of background uncertainties requires the experiment-level likelihood and is beyond the scope of this work.

\section{Results and discussion \label{sec:ExpectedReach}}

In this section, we discuss the direct-detection constraints on light inelastic DM (see 
Eq.~\eqref{eq:EffLagrangianScalarMEDMajoranaiCDM}) with a scalar leptophilic mediator, within the parameter space 
defined in Eq.~\eqref{eq:paramSpace}.
Specifically, the PandaX-4T, XENON1T, and LZ experiments considered here are sensitive to deposited energies in 
the signal region of order $\mathcal{O}(10^{-8})~\mbox{GeV}$, $\mathcal{O}(10^{-7})~\mbox{GeV}$, and 
$\mathcal{O}(10^{-6})~\mbox{GeV}$, respectively.
We employ both binned Eq.~\eqref{eq:binnedStatistics} and unbinned Eq.~\eqref{eq:unBinnedStatistics}
approaches to estimate the direct-detection constraints.
It is also important to note that the limit obtained using the unbinned 
approach  Eq.~\eqref{eq:unBinnedStatistics} allows one to derive direct-detection constraints up to an additional 
$\mathcal{O}(1)$ factor compared to the likelihood-based estimate, for the direct-detection experiments 
considered here. The corresponding direct-detection constraints for the endothermic and exothermic scenarios are 
shown in Figs.~\ref{fig:DDconstraintUpScattering} and~\ref{fig:DDconstraintDownScattering}, respectively.

In order to map the constraints from the BaBar and NA62 experiments into parameter space 
of interest, we employed the reference cross section~\eqref{eq:ExpliсMajoranaDMCS} where 
the corresponding dependence $c_{ee}^{\phi}(m_{\phi})$ was taken from the 
Ref.~\cite{BaBar:2017tiz} and Ref.~\cite{Krnjaic:2019rsv,NA62:2021bji}, respectively. 
Similarly, we use the constraints on scalar-mediator radiation from fixed-target 
experiments~\cite{Voronchikhin:2025eqm}.
We also compare our computed results in the limit of zero mass splitting with known 
results of XENON1T and PandaX-4T experiments, and the resulting curves agree at the level 
of an $\mathcal{O}(1)$ factor.
We also use the relic-density curves obtained in our previous work~\cite{Voronchikhin:2025eqm}, where the mass splitting has no significant impact on the freeze-out mechanism in the considered parameter region~\eqref{eq:paramSpace}.
Note that both NA64e and  NA62 experiments rule out the typical  masses~$m_{\chi_1}~\lesssim~100~\mbox{MeV}$ within the adopted thermal target benchmarks. 

\textbf{Endothermic scenario (up-scattering, Fig.~\ref{fig:DDconstraintUpScattering})}. 
In the case of an endothermic reaction, increasing the mass splitting leads to weaker 
constraints, up to the kinematically forbidden region $\Delta > 10^{-6}$. 
However, for the relative mass splittings $\Delta = 10^{-7}$, the constraints for up-scattering 
process differ from the elastic case only at the level of an $\mathcal{O}(1)$ factor.
Moreover, for smaller mass splittings, $\Delta \ll 10^{-7} $, the 
constraints on inelastic DM are close to the elastic case.
Indeed, as the mass splitting between the states decreases, its impact on the minimum 
dark-matter velocity, Eq.~(\ref{eq:VminDef}), becomes smaller for an endothermic reaction. 
 For the $\Delta~\simeq~10^{-7}$, significant deviations from the elastic  
case $\Delta  = 0$ appear at masses $m_{\chi_1}~> 1~\mbox{GeV}$ for LZ and a 
$m_{\chi_1}~> 100~\mbox{MeV}$ for XENON1T and PandaX-4T.
Thus, sufficiently small mass splittings $\Delta \ll 10^{-7}$ do not lead 
to any additional weakening of the direct-detection constraints.  

\textbf{Exothermic scenario (down-scattering, Fig.~\ref{fig:DDconstraintDownScattering})}.
In the case of exothermic scattering, the typical masses about $m_{\chi_1}~\simeq~E_{\rm d}/|\Delta|$ provide better sensitivity (this corresponds to the lowest upper limit on $\bar{\sigma}_e$ shown in Fig.~\ref{fig:DDconstraintDownScattering}), for which the lower bound of dark matter velocity is minimal, see Eq.~(\ref{eq:VminDef}). 
Given $\Delta$, the corresponding sensitivity enhancement arises from the effect of the 
mass splitting on the kinematic quantities, as can be seen directly from 
Figs.~\ref{fig:qMinMax} and~\ref{fig:IntgrndDiffRate} (see Sec.~\ref{subsect:Kinem} for a 
more detailed discussion). In particular, for a relative mass splitting of $|\Delta| = 10^{-5}$, the sensitivity peaks arise near DM masses of $\mathcal{O}(10^{-2})\ 
\mbox{GeV}$, $\mathcal{O}(10^{-1})\ \mbox{GeV}$, and $\mathcal{O}(1)\ \mbox{GeV}$ for 
PandaX-4T, XENON1T, and LZ, respectively.

The constraints at relatively large masses, $m_{\chi_1} \gtrsim E_{\rm d}/|\Delta|$, are weakened compared to the peak sensitivity region $m_{\chi_1} \simeq E_{\rm d}/|\Delta|$ due to kinematic suppression, which implies shrinking of the integration limits Eq.~\eqref{eq:rangeTransfMomenta}.
At relatively small masses, $m_{\chi_1}~\lesssim~E_{\rm d}/|\Delta|$, the sensitivities shift toward larger 
cross sections. 
 As the relative mass splitting increases, direct-detection experiments rule out sufficiently large portion of the parameter space, within small DM mass region, 
 i.~e.~the larger $|\Delta|$, the better the sensitivity. However, sufficiently large $|\Delta| \gtrsim 10^{-5}$ are  forbidden due to kinematics.  

It is worth noticing, the bounds for exothermic scenario, shown in Fig.~\ref{fig:DDconstraintDownScattering}, are 
sensitive to the energy binned distribution,  this leads to irregularities in the sensitivity curves.
Remarkable, the extended limits from considered direct-detection 
experiments are comparable to expected from the NA64$\mu$ experiment 
with a statistics of~$10^{11}$ muons on target, which is the most 
sensitive probe of this model.

\section{Conclusion\label{sec:Conclusion}}

In this work, we derived direct-detection constraints on thermal 
inelastic fermion dark matter coupled to a leptophilic scalar mediator 
using public data from XENON1T, PandaX-4T, and LZ.
The mass splitting $|\Delta| \simeq \mathcal{O}(10^{-5})-\mathcal{O}(10^{-6})$ for 
the down-scattering setup can lead to a enhancement of the sensitivity for 
the experimental facilities of interest.
For a heavy-state fraction $f_{\chi_2} = 1/2$ and the exothermic scenario, the 
signal can be sufficiently enhanced in the characteristic mass regions, $m_{\chi_1}~\simeq~E_{\rm d}/|\Delta|$, up to the 
level $\sigma_e \simeq \mathcal{O}(10^{-45})~\mbox{cm}^2$.
In this case, direct-detection experiments can additionally exclude the parameter 
space in the mass range, 
$100~\mbox{MeV} < m_{\chi_1} < 500~\mbox{MeV}$. However,
following Eq.~(\ref{eq:dRdE}), we note that a smaller fraction $f_{\chi_2} \ll 1/2$ 
would result in a weaker constraint.

\begin{acknowledgments} 
The work on deriving sensitivities for XENON1T and PandaX-4T  was supported by the Foundation for the Advancement of Theoretical 
Physics and Mathematics BASIS (Project No.~\text{24-1-2-11-2} and No.~\text{24-1-2-11-1}).
The  work on deriving sensitivities for LZ was  
supported by the the Russian Science Foundation grant No.~25-12-00309.
\end{acknowledgments}

\appendix

\bibliography{bibl}

\begin{thebibliography}{98}%
\makeatletter
\providecommand \@ifxundefined [1]{%
 \@ifx{#1\undefined}
}%
\providecommand \@ifnum [1]{%
 \ifnum #1\expandafter \@firstoftwo
 \else \expandafter \@secondoftwo
 \fi
}%
\providecommand \@ifx [1]{%
 \ifx #1\expandafter \@firstoftwo
 \else \expandafter \@secondoftwo
 \fi
}%
\providecommand \natexlab [1]{#1}%
\providecommand \enquote  [1]{``#1''}%
\providecommand \bibnamefont  [1]{#1}%
\providecommand \bibfnamefont [1]{#1}%
\providecommand \citenamefont [1]{#1}%
\providecommand \href@noop [0]{\@secondoftwo}%
\providecommand \href [0]{\begingroup \@sanitize@url \@href}%
\providecommand \@href[1]{\@@startlink{#1}\@@href}%
\providecommand \@@href[1]{\endgroup#1\@@endlink}%
\providecommand \@sanitize@url [0]{\catcode `\\12\catcode `\$12\catcode `\&12\catcode `\#12\catcode `\^12\catcode `\_12\catcode `\%12\relax}%
\providecommand \@@startlink[1]{}%
\providecommand \@@endlink[0]{}%
\providecommand \url  [0]{\begingroup\@sanitize@url \@url }%
\providecommand \@url [1]{\endgroup\@href {#1}{\urlprefix }}%
\providecommand \urlprefix  [0]{URL }%
\providecommand \Eprint [0]{\href }%
\providecommand \doibase [0]{http://dx.doi.org/}%
\providecommand \selectlanguage [0]{\@gobble}%
\providecommand \bibinfo  [0]{\@secondoftwo}%
\providecommand \bibfield  [0]{\@secondoftwo}%
\providecommand \translation [1]{[#1]}%
\providecommand \BibitemOpen [0]{}%
\providecommand \bibitemStop [0]{}%
\providecommand \bibitemNoStop [0]{.\EOS\space}%
\providecommand \EOS [0]{\spacefactor3000\relax}%
\providecommand \BibitemShut  [1]{\csname bibitem#1\endcsname}%
\let\auto@bib@innerbib\@empty
\bibitem [{\citenamefont {Aghanim}\ \emph {et~al.}(2020)\citenamefont {Aghanim} \emph {et~al.}}]{Planck:2018vyg}%
  \BibitemOpen
  \bibfield  {author} {\bibinfo {author} {\bibfnamefont {N.}~\bibnamefont {Aghanim}} \emph {et~al.} (\bibinfo {collaboration} {Planck}),\ }\bibfield  {title} {\enquote {\bibinfo {title} {{Planck 2018 results. VI. Cosmological parameters}},}\ }\href {\doibase 10.1051/0004-6361/201833910} {\bibfield  {journal} {\bibinfo  {journal} {Astron. Astrophys.}\ }\textbf {\bibinfo {volume} {641}},\ \bibinfo {pages} {A6} (\bibinfo {year} {2020})},\ \bibinfo {note} {[Erratum: Astron.Astrophys. 652, C4 (2021)]},\ \Eprint {http://arxiv.org/abs/1807.06209} {arXiv:1807.06209 [astro-ph.CO]} \BibitemShut {NoStop}%
\bibitem [{\citenamefont {Bertone}\ and\ \citenamefont {Hooper}(2018)}]{Bertone:2016nfn}%
  \BibitemOpen
  \bibfield  {author} {\bibinfo {author} {\bibfnamefont {Gianfranco}\ \bibnamefont {Bertone}}\ and\ \bibinfo {author} {\bibfnamefont {Dan}\ \bibnamefont {Hooper}},\ }\bibfield  {title} {\enquote {\bibinfo {title} {{History of dark matter}},}\ }\href {\doibase 10.1103/RevModPhys.90.045002} {\bibfield  {journal} {\bibinfo  {journal} {Rev. Mod. Phys.}\ }\textbf {\bibinfo {volume} {90}},\ \bibinfo {pages} {045002} (\bibinfo {year} {2018})},\ \Eprint {http://arxiv.org/abs/1605.04909} {arXiv:1605.04909 [astro-ph.CO]} \BibitemShut {NoStop}%
\bibitem [{\citenamefont {Aprile}\ \emph {et~al.}(2018)\citenamefont {Aprile} \emph {et~al.}}]{XENON:2018voc}%
  \BibitemOpen
  \bibfield  {author} {\bibinfo {author} {\bibfnamefont {E.}~\bibnamefont {Aprile}} \emph {et~al.} (\bibinfo {collaboration} {XENON}),\ }\bibfield  {title} {\enquote {\bibinfo {title} {{Dark Matter Search Results from a One Ton-Year Exposure of XENON1T}},}\ }\href {\doibase 10.1103/PhysRevLett.121.111302} {\bibfield  {journal} {\bibinfo  {journal} {Phys. Rev. Lett.}\ }\textbf {\bibinfo {volume} {121}},\ \bibinfo {pages} {111302} (\bibinfo {year} {2018})},\ \Eprint {http://arxiv.org/abs/1805.12562} {arXiv:1805.12562 [astro-ph.CO]} \BibitemShut {NoStop}%
\bibitem [{\citenamefont {Meng}\ \emph {et~al.}(2021)\citenamefont {Meng} \emph {et~al.}}]{PandaX-4T:2021bab}%
  \BibitemOpen
  \bibfield  {author} {\bibinfo {author} {\bibfnamefont {Yue}\ \bibnamefont {Meng}} \emph {et~al.} (\bibinfo {collaboration} {PandaX-4T}),\ }\bibfield  {title} {\enquote {\bibinfo {title} {{Dark Matter Search Results from the PandaX-4T Commissioning Run}},}\ }\href {\doibase 10.1103/PhysRevLett.127.261802} {\bibfield  {journal} {\bibinfo  {journal} {Phys. Rev. Lett.}\ }\textbf {\bibinfo {volume} {127}},\ \bibinfo {pages} {261802} (\bibinfo {year} {2021})},\ \Eprint {http://arxiv.org/abs/2107.13438} {arXiv:2107.13438 [hep-ex]} \BibitemShut {NoStop}%
\bibitem [{\citenamefont {Aalbers}\ \emph {et~al.}(2023)\citenamefont {Aalbers} \emph {et~al.}}]{LZ:2022lsv}%
  \BibitemOpen
  \bibfield  {author} {\bibinfo {author} {\bibfnamefont {J.}~\bibnamefont {Aalbers}} \emph {et~al.} (\bibinfo {collaboration} {LZ}),\ }\bibfield  {title} {\enquote {\bibinfo {title} {{First Dark Matter Search Results from the LUX-ZEPLIN (LZ) Experiment}},}\ }\href {\doibase 10.1103/PhysRevLett.131.041002} {\bibfield  {journal} {\bibinfo  {journal} {Phys. Rev. Lett.}\ }\textbf {\bibinfo {volume} {131}},\ \bibinfo {pages} {041002} (\bibinfo {year} {2023})},\ \Eprint {http://arxiv.org/abs/2207.03764} {arXiv:2207.03764 [hep-ex]} \BibitemShut {NoStop}%
\bibitem [{\citenamefont {Lee}\ and\ \citenamefont {Weinberg}(1977)}]{Lee:1977ua}%
  \BibitemOpen
  \bibfield  {author} {\bibinfo {author} {\bibfnamefont {Benjamin~W.}\ \bibnamefont {Lee}}\ and\ \bibinfo {author} {\bibfnamefont {Steven}\ \bibnamefont {Weinberg}},\ }\bibfield  {title} {\enquote {\bibinfo {title} {{Cosmological Lower Bound on Heavy Neutrino Masses}},}\ }\href {\doibase 10.1103/PhysRevLett.39.165} {\bibfield  {journal} {\bibinfo  {journal} {Phys. Rev. Lett.}\ }\textbf {\bibinfo {volume} {39}},\ \bibinfo {pages} {165--168} (\bibinfo {year} {1977})}\BibitemShut {NoStop}%
\bibitem [{\citenamefont {Kolb}\ and\ \citenamefont {Olive}(1986)}]{Kolb:1985nn}%
  \BibitemOpen
  \bibfield  {author} {\bibinfo {author} {\bibfnamefont {Edward~W.}\ \bibnamefont {Kolb}}\ and\ \bibinfo {author} {\bibfnamefont {Keith~A.}\ \bibnamefont {Olive}},\ }\bibfield  {title} {\enquote {\bibinfo {title} {{The Lee-Weinberg Bound Revisited}},}\ }\href {\doibase 10.1103/PhysRevD.33.1202} {\bibfield  {journal} {\bibinfo  {journal} {Phys. Rev. D}\ }\textbf {\bibinfo {volume} {33}},\ \bibinfo {pages} {1202} (\bibinfo {year} {1986})},\ \bibinfo {note} {[Erratum: Phys.Rev.D 34, 2531 (1986)]}\BibitemShut {NoStop}%
\bibitem [{\citenamefont {Krnjaic}(2016)}]{Krnjaic:2015mbs}%
  \BibitemOpen
  \bibfield  {author} {\bibinfo {author} {\bibfnamefont {Gordan}\ \bibnamefont {Krnjaic}},\ }\bibfield  {title} {\enquote {\bibinfo {title} {{Probing Light Thermal Dark-Matter With a Higgs Portal Mediator}},}\ }\href {\doibase 10.1103/PhysRevD.94.073009} {\bibfield  {journal} {\bibinfo  {journal} {Phys. Rev. D}\ }\textbf {\bibinfo {volume} {94}},\ \bibinfo {pages} {073009} (\bibinfo {year} {2016})},\ \Eprint {http://arxiv.org/abs/1512.04119} {arXiv:1512.04119 [hep-ph]} \BibitemShut {NoStop}%
\bibitem [{\citenamefont {McDonald}(1994)}]{McDonald:1993ex}%
  \BibitemOpen
  \bibfield  {author} {\bibinfo {author} {\bibfnamefont {John}\ \bibnamefont {McDonald}},\ }\bibfield  {title} {\enquote {\bibinfo {title} {{Gauge singlet scalars as cold dark matter}},}\ }\href {\doibase 10.1103/PhysRevD.50.3637} {\bibfield  {journal} {\bibinfo  {journal} {Phys. Rev. D}\ }\textbf {\bibinfo {volume} {50}},\ \bibinfo {pages} {3637--3649} (\bibinfo {year} {1994})},\ \Eprint {http://arxiv.org/abs/hep-ph/0702143} {arXiv:hep-ph/0702143} \BibitemShut {NoStop}%
\bibitem [{\citenamefont {Burgess}\ \emph {et~al.}(2001)\citenamefont {Burgess}, \citenamefont {Pospelov},\ and\ \citenamefont {ter Veldhuis}}]{Burgess:2000yq}%
  \BibitemOpen
  \bibfield  {author} {\bibinfo {author} {\bibfnamefont {C.~P.}\ \bibnamefont {Burgess}}, \bibinfo {author} {\bibfnamefont {Maxim}\ \bibnamefont {Pospelov}}, \ and\ \bibinfo {author} {\bibfnamefont {Tonnis}\ \bibnamefont {ter Veldhuis}},\ }\bibfield  {title} {\enquote {\bibinfo {title} {{The Minimal model of nonbaryonic dark matter: A Singlet scalar}},}\ }\href {\doibase 10.1016/S0550-3213(01)00513-2} {\bibfield  {journal} {\bibinfo  {journal} {Nucl. Phys. B}\ }\textbf {\bibinfo {volume} {619}},\ \bibinfo {pages} {709--728} (\bibinfo {year} {2001})},\ \Eprint {http://arxiv.org/abs/hep-ph/0011335} {arXiv:hep-ph/0011335} \BibitemShut {NoStop}%
\bibitem [{\citenamefont {Wells}(2008)}]{Wells:2008xg}%
  \BibitemOpen
  \bibfield  {author} {\bibinfo {author} {\bibfnamefont {James~D.}\ \bibnamefont {Wells}},\ }\bibfield  {title} {\enquote {\bibinfo {title} {{How to Find a Hidden World at the Large Hadron Collider}},}\ }\href@noop {} {\ ,\ \bibinfo {pages} {283--298} (\bibinfo {year} {2008})},\ \Eprint {http://arxiv.org/abs/0803.1243} {arXiv:0803.1243 [hep-ph]} \BibitemShut {NoStop}%
\bibitem [{\citenamefont {Sieber}\ \emph {et~al.}(2023)\citenamefont {Sieber}, \citenamefont {Kirpichnikov}, \citenamefont {Voronchikhin}, \citenamefont {Crivelli}, \citenamefont {Gninenko}, \citenamefont {Kirsanov}, \citenamefont {Krasnikov}, \citenamefont {Molina-Bueno},\ and\ \citenamefont {Sekatskii}}]{Sieber:2023nkq}%
  \BibitemOpen
  \bibfield  {author} {\bibinfo {author} {\bibfnamefont {H.}~\bibnamefont {Sieber}}, \bibinfo {author} {\bibfnamefont {D.~V.}\ \bibnamefont {Kirpichnikov}}, \bibinfo {author} {\bibfnamefont {I.~V.}\ \bibnamefont {Voronchikhin}}, \bibinfo {author} {\bibfnamefont {P.}~\bibnamefont {Crivelli}}, \bibinfo {author} {\bibfnamefont {S.~N.}\ \bibnamefont {Gninenko}}, \bibinfo {author} {\bibfnamefont {M.~M.}\ \bibnamefont {Kirsanov}}, \bibinfo {author} {\bibfnamefont {N.~V.}\ \bibnamefont {Krasnikov}}, \bibinfo {author} {\bibfnamefont {L.}~\bibnamefont {Molina-Bueno}}, \ and\ \bibinfo {author} {\bibfnamefont {S.~K.}\ \bibnamefont {Sekatskii}},\ }\bibfield  {title} {\enquote {\bibinfo {title} {{Probing hidden sectors with a muon beam: Implication of spin-0 dark matter mediators for the muon (g-2) anomaly and the validity of the Weisz{\"a}cker-Williams approach}},}\ }\href {\doibase 10.1103/PhysRevD.108.056018} {\bibfield  {journal} {\bibinfo  {journal} {Phys. Rev. D}\ }\textbf {\bibinfo {volume} {108}},\ \bibinfo
  {pages} {056018} (\bibinfo {year} {2023})},\ \Eprint {http://arxiv.org/abs/2305.09015} {arXiv:2305.09015 [hep-ph]} \BibitemShut {NoStop}%
\bibitem [{\citenamefont {Guo}\ \emph {et~al.}(2025)\citenamefont {Guo}, \citenamefont {Liu}, \citenamefont {Peng},\ and\ \citenamefont {Wang}}]{Guo:2025qes}%
  \BibitemOpen
  \bibfield  {author} {\bibinfo {author} {\bibfnamefont {Jinhui}\ \bibnamefont {Guo}}, \bibinfo {author} {\bibfnamefont {Jia}\ \bibnamefont {Liu}}, \bibinfo {author} {\bibfnamefont {Chenhao}\ \bibnamefont {Peng}}, \ and\ \bibinfo {author} {\bibfnamefont {Xiao-Ping}\ \bibnamefont {Wang}},\ }\bibfield  {title} {\enquote {\bibinfo {title} {{Probing purely inelastic scalar dark matter across colliders and gravitational wave observatories}},}\ }\href {\doibase 10.1103/9myr-9xx5} {\bibfield  {journal} {\bibinfo  {journal} {Phys. Rev. D}\ }\textbf {\bibinfo {volume} {112}},\ \bibinfo {pages} {115014} (\bibinfo {year} {2025})},\ \Eprint {http://arxiv.org/abs/2508.13276} {arXiv:2508.13276 [hep-ph]} \BibitemShut {NoStop}%
\bibitem [{\citenamefont {Voronchikhin}\ and\ \citenamefont {Kirpichnikov}(2024)}]{Voronchikhin:2023qig}%
  \BibitemOpen
  \bibfield  {author} {\bibinfo {author} {\bibfnamefont {I.~V.}\ \bibnamefont {Voronchikhin}}\ and\ \bibinfo {author} {\bibfnamefont {D.~V.}\ \bibnamefont {Kirpichnikov}},\ }\bibfield  {title} {\enquote {\bibinfo {title} {{Probing scalar, Dirac, Majorana, and vector dark matter through a spin-0 electron-specific mediator at electron fixed-target experiments}},}\ }\href {\doibase 10.1103/PhysRevD.109.075012} {\bibfield  {journal} {\bibinfo  {journal} {Phys. Rev. D}\ }\textbf {\bibinfo {volume} {109}},\ \bibinfo {pages} {075012} (\bibinfo {year} {2024})},\ \Eprint {http://arxiv.org/abs/2312.15697} {arXiv:2312.15697 [hep-ph]} \BibitemShut {NoStop}%
\bibitem [{\citenamefont {Becker}\ \emph {et~al.}(2025)\citenamefont {Becker}, \citenamefont {Copello}, \citenamefont {Harz},\ and\ \citenamefont {Napetschnig}}]{Becker:2025vgq}%
  \BibitemOpen
  \bibfield  {author} {\bibinfo {author} {\bibfnamefont {Mathias}\ \bibnamefont {Becker}}, \bibinfo {author} {\bibfnamefont {Emanuele}\ \bibnamefont {Copello}}, \bibinfo {author} {\bibfnamefont {Julia}\ \bibnamefont {Harz}}, \ and\ \bibinfo {author} {\bibfnamefont {Martin}\ \bibnamefont {Napetschnig}},\ }\bibfield  {title} {\enquote {\bibinfo {title} {{Manual for SE+BSF4DM -- A micrOMEGAs package for Sommerfeld Effect and Bound State Formation in colored Dark Sectors}},}\ }\href@noop {} {\  (\bibinfo {year} {2025})},\ \Eprint {http://arxiv.org/abs/2512.02155} {arXiv:2512.02155 [hep-ph]} \BibitemShut {NoStop}%
\bibitem [{\citenamefont {Becker}\ \emph {et~al.}(2026)\citenamefont {Becker}, \citenamefont {Copello}, \citenamefont {Harz},\ and\ \citenamefont {Napetschnig}}]{Becker:2026icc}%
  \BibitemOpen
  \bibfield  {author} {\bibinfo {author} {\bibfnamefont {Mathias}\ \bibnamefont {Becker}}, \bibinfo {author} {\bibfnamefont {Emanuele}\ \bibnamefont {Copello}}, \bibinfo {author} {\bibfnamefont {Julia}\ \bibnamefont {Harz}}, \ and\ \bibinfo {author} {\bibfnamefont {Martin}\ \bibnamefont {Napetschnig}},\ }\bibfield  {title} {\enquote {\bibinfo {title} {{Sommerfeld Effect and Bound State Formation for Dark Matter Models with Colored Mediators with SE+BSF4DM}},}\ }\href@noop {} {\  (\bibinfo {year} {2026})},\ \Eprint {http://arxiv.org/abs/2601.03026} {arXiv:2601.03026 [hep-ph]} \BibitemShut {NoStop}%
\bibitem [{\citenamefont {Wang}(2025)}]{Wang:2025xoq}%
  \BibitemOpen
  \bibfield  {author} {\bibinfo {author} {\bibfnamefont {Zihan}\ \bibnamefont {Wang}},\ }\bibfield  {title} {\enquote {\bibinfo {title} {{Scalar-Mediated Inelastic Dark Matter as a Solution to Small-Scale Structure Anomalies}},}\ }\href@noop {} {\  (\bibinfo {year} {2025})},\ \Eprint {http://arxiv.org/abs/2512.18959} {arXiv:2512.18959 [hep-ph]} \BibitemShut {NoStop}%
\bibitem [{\citenamefont {Holdom}(1986)}]{Holdom:1985ag}%
  \BibitemOpen
  \bibfield  {author} {\bibinfo {author} {\bibfnamefont {Bob}\ \bibnamefont {Holdom}},\ }\bibfield  {title} {\enquote {\bibinfo {title} {{Two U(1)'s and Epsilon Charge Shifts}},}\ }\href {\doibase 10.1016/0370-2693(86)91377-8} {\bibfield  {journal} {\bibinfo  {journal} {Phys. Lett. B}\ }\textbf {\bibinfo {volume} {166}},\ \bibinfo {pages} {196--198} (\bibinfo {year} {1986})}\BibitemShut {NoStop}%
\bibitem [{\citenamefont {Izaguirre}\ \emph {et~al.}(2015)\citenamefont {Izaguirre}, \citenamefont {Krnjaic}, \citenamefont {Schuster},\ and\ \citenamefont {Toro}}]{Izaguirre:2015yja}%
  \BibitemOpen
  \bibfield  {author} {\bibinfo {author} {\bibfnamefont {Eder}\ \bibnamefont {Izaguirre}}, \bibinfo {author} {\bibfnamefont {Gordan}\ \bibnamefont {Krnjaic}}, \bibinfo {author} {\bibfnamefont {Philip}\ \bibnamefont {Schuster}}, \ and\ \bibinfo {author} {\bibfnamefont {Natalia}\ \bibnamefont {Toro}},\ }\bibfield  {title} {\enquote {\bibinfo {title} {{Analyzing the Discovery Potential for Light Dark Matter}},}\ }\href {\doibase 10.1103/PhysRevLett.115.251301} {\bibfield  {journal} {\bibinfo  {journal} {Phys. Rev. Lett.}\ }\textbf {\bibinfo {volume} {115}},\ \bibinfo {pages} {251301} (\bibinfo {year} {2015})},\ \Eprint {http://arxiv.org/abs/1505.00011} {arXiv:1505.00011 [hep-ph]} \BibitemShut {NoStop}%
\bibitem [{\citenamefont {Batell}\ \emph {et~al.}(2014)\citenamefont {Batell}, \citenamefont {Essig},\ and\ \citenamefont {Surujon}}]{Batell:2014mga}%
  \BibitemOpen
  \bibfield  {author} {\bibinfo {author} {\bibfnamefont {Brian}\ \bibnamefont {Batell}}, \bibinfo {author} {\bibfnamefont {Rouven}\ \bibnamefont {Essig}}, \ and\ \bibinfo {author} {\bibfnamefont {Ze'ev}\ \bibnamefont {Surujon}},\ }\bibfield  {title} {\enquote {\bibinfo {title} {{Strong Constraints on Sub-GeV Dark Sectors from SLAC Beam Dump E137}},}\ }\href {\doibase 10.1103/PhysRevLett.113.171802} {\bibfield  {journal} {\bibinfo  {journal} {Phys. Rev. Lett.}\ }\textbf {\bibinfo {volume} {113}},\ \bibinfo {pages} {171802} (\bibinfo {year} {2014})},\ \Eprint {http://arxiv.org/abs/1406.2698} {arXiv:1406.2698 [hep-ph]} \BibitemShut {NoStop}%
\bibitem [{\citenamefont {Kachanovich}\ \emph {et~al.}(2022)\citenamefont {Kachanovich}, \citenamefont {Kovalenko}, \citenamefont {Kuleshov}, \citenamefont {Lyubovitskij},\ and\ \citenamefont {Zhevlakov}}]{Kachanovich:2021eqa}%
  \BibitemOpen
  \bibfield  {author} {\bibinfo {author} {\bibfnamefont {Aliaksei}\ \bibnamefont {Kachanovich}}, \bibinfo {author} {\bibfnamefont {Sergey}\ \bibnamefont {Kovalenko}}, \bibinfo {author} {\bibfnamefont {Serguei}\ \bibnamefont {Kuleshov}}, \bibinfo {author} {\bibfnamefont {Valery~E.}\ \bibnamefont {Lyubovitskij}}, \ and\ \bibinfo {author} {\bibfnamefont {Alexey~S.}\ \bibnamefont {Zhevlakov}},\ }\bibfield  {title} {\enquote {\bibinfo {title} {{Lepton phenomenology of Stueckelberg portal to dark sector}},}\ }\href {\doibase 10.1103/PhysRevD.105.075004} {\bibfield  {journal} {\bibinfo  {journal} {Phys. Rev. D}\ }\textbf {\bibinfo {volume} {105}},\ \bibinfo {pages} {075004} (\bibinfo {year} {2022})},\ \Eprint {http://arxiv.org/abs/2111.12522} {arXiv:2111.12522 [hep-ph]} \BibitemShut {NoStop}%
\bibitem [{\citenamefont {Lyubovitskij}\ \emph {et~al.}(2023)\citenamefont {Lyubovitskij}, \citenamefont {Zhevlakov}, \citenamefont {Kachanovich},\ and\ \citenamefont {Kuleshov}}]{Lyubovitskij:2022hna}%
  \BibitemOpen
  \bibfield  {author} {\bibinfo {author} {\bibfnamefont {Valery~E.}\ \bibnamefont {Lyubovitskij}}, \bibinfo {author} {\bibfnamefont {Alexey~S.}\ \bibnamefont {Zhevlakov}}, \bibinfo {author} {\bibfnamefont {Aliaksei}\ \bibnamefont {Kachanovich}}, \ and\ \bibinfo {author} {\bibfnamefont {Serguei}\ \bibnamefont {Kuleshov}},\ }\bibfield  {title} {\enquote {\bibinfo {title} {{Dark $SU(2)$ Stueckelberg portal}},}\ }\href {\doibase 10.1103/PhysRevD.107.055006} {\bibfield  {journal} {\bibinfo  {journal} {Phys. Rev. D}\ }\textbf {\bibinfo {volume} {107}},\ \bibinfo {pages} {055006} (\bibinfo {year} {2023})},\ \Eprint {http://arxiv.org/abs/2210.05555} {arXiv:2210.05555 [hep-ph]} \BibitemShut {NoStop}%
\bibitem [{\citenamefont {Gorbunov}\ and\ \citenamefont {Kalashnikov}(2023)}]{Gorbunov:2022dgw}%
  \BibitemOpen
  \bibfield  {author} {\bibinfo {author} {\bibfnamefont {Dmitry}\ \bibnamefont {Gorbunov}}\ and\ \bibinfo {author} {\bibfnamefont {Dmitry}\ \bibnamefont {Kalashnikov}},\ }\bibfield  {title} {\enquote {\bibinfo {title} {{Probing light exotics from a hidden sector at c-{\ensuremath{\tau}} factories with polarized electron beams}},}\ }\href {\doibase 10.1103/PhysRevD.107.015014} {\bibfield  {journal} {\bibinfo  {journal} {Phys. Rev. D}\ }\textbf {\bibinfo {volume} {107}},\ \bibinfo {pages} {015014} (\bibinfo {year} {2023})},\ \Eprint {http://arxiv.org/abs/2211.06270} {arXiv:2211.06270 [hep-ph]} \BibitemShut {NoStop}%
\bibitem [{\citenamefont {Claude}\ \emph {et~al.}(2023)\citenamefont {Claude}, \citenamefont {Dutra},\ and\ \citenamefont {Godfrey}}]{Claude:2022rho}%
  \BibitemOpen
  \bibfield  {author} {\bibinfo {author} {\bibfnamefont {J{\'e}r{\^o}me}\ \bibnamefont {Claude}}, \bibinfo {author} {\bibfnamefont {Ma{\'\i}ra}\ \bibnamefont {Dutra}}, \ and\ \bibinfo {author} {\bibfnamefont {Stephen}\ \bibnamefont {Godfrey}},\ }\bibfield  {title} {\enquote {\bibinfo {title} {{Probing feebly interacting dark matter with monojet searches}},}\ }\href {\doibase 10.1103/PhysRevD.107.075006} {\bibfield  {journal} {\bibinfo  {journal} {Phys. Rev. D}\ }\textbf {\bibinfo {volume} {107}},\ \bibinfo {pages} {075006} (\bibinfo {year} {2023})},\ \Eprint {http://arxiv.org/abs/2208.09422} {arXiv:2208.09422 [hep-ph]} \BibitemShut {NoStop}%
\bibitem [{\citenamefont {Wang}\ \emph {et~al.}(2023)\citenamefont {Wang}, \citenamefont {Xu}, \citenamefont {Yang},\ and\ \citenamefont {Zhu}}]{Wang:2023wrx}%
  \BibitemOpen
  \bibfield  {author} {\bibinfo {author} {\bibfnamefont {Wenyu}\ \bibnamefont {Wang}}, \bibinfo {author} {\bibfnamefont {Wu-Long}\ \bibnamefont {Xu}}, \bibinfo {author} {\bibfnamefont {Jin~Min}\ \bibnamefont {Yang}}, \ and\ \bibinfo {author} {\bibfnamefont {Rui}\ \bibnamefont {Zhu}},\ }\bibfield  {title} {\enquote {\bibinfo {title} {{Direct detection of cosmic ray-boosted puffy dark matter}},}\ }\href {\doibase 10.1016/j.nuclphysb.2023.116348} {\bibfield  {journal} {\bibinfo  {journal} {Nucl. Phys. B}\ }\textbf {\bibinfo {volume} {995}},\ \bibinfo {pages} {116348} (\bibinfo {year} {2023})},\ \Eprint {http://arxiv.org/abs/2305.12668} {arXiv:2305.12668 [hep-ph]} \BibitemShut {NoStop}%
\bibitem [{\citenamefont {Voronchikhin}\ and\ \citenamefont {Kirpichnikov}(2025{\natexlab{a}})}]{Voronchikhin:2024vfu}%
  \BibitemOpen
  \bibfield  {author} {\bibinfo {author} {\bibfnamefont {I.~V.}\ \bibnamefont {Voronchikhin}}\ and\ \bibinfo {author} {\bibfnamefont {D.~V.}\ \bibnamefont {Kirpichnikov}},\ }\bibfield  {title} {\enquote {\bibinfo {title} {{Implication of the Weizsacker-Williams approximation for the dark matter mediator production}},}\ }\href {\doibase 10.1103/PhysRevD.111.035034} {\bibfield  {journal} {\bibinfo  {journal} {Phys. Rev. D}\ }\textbf {\bibinfo {volume} {111}},\ \bibinfo {pages} {035034} (\bibinfo {year} {2025}{\natexlab{a}})},\ \Eprint {http://arxiv.org/abs/2409.12748} {arXiv:2409.12748 [hep-ph]} \BibitemShut {NoStop}%
\bibitem [{\citenamefont {Gustafson}\ \emph {et~al.}(2025{\natexlab{a}})\citenamefont {Gustafson}, \citenamefont {Herrera}, \citenamefont {Mukhopadhyay}, \citenamefont {Murase},\ and\ \citenamefont {Shoemaker}}]{Gustafson:2025dff}%
  \BibitemOpen
  \bibfield  {author} {\bibinfo {author} {\bibfnamefont {R.~Andrew}\ \bibnamefont {Gustafson}}, \bibinfo {author} {\bibfnamefont {Gonzalo}\ \bibnamefont {Herrera}}, \bibinfo {author} {\bibfnamefont {Mainak}\ \bibnamefont {Mukhopadhyay}}, \bibinfo {author} {\bibfnamefont {Kohta}\ \bibnamefont {Murase}}, \ and\ \bibinfo {author} {\bibfnamefont {Ian~M.}\ \bibnamefont {Shoemaker}},\ }\bibfield  {title} {\enquote {\bibinfo {title} {{Cosmic-ray boosted inelastic dark matter from neutrino-emitting active galactic nuclei}},}\ }\href@noop {} {\  (\bibinfo {year} {2025}{\natexlab{a}})},\ \Eprint {http://arxiv.org/abs/2508.20984} {arXiv:2508.20984 [hep-ph]} \BibitemShut {NoStop}%
\bibitem [{\citenamefont {Herrera}\ \emph {et~al.}(2023)\citenamefont {Herrera}, \citenamefont {Ibarra},\ and\ \citenamefont {Shirai}}]{Herrera:2023fpq}%
  \BibitemOpen
  \bibfield  {author} {\bibinfo {author} {\bibfnamefont {Gonzalo}\ \bibnamefont {Herrera}}, \bibinfo {author} {\bibfnamefont {Alejandro}\ \bibnamefont {Ibarra}}, \ and\ \bibinfo {author} {\bibfnamefont {Satoshi}\ \bibnamefont {Shirai}},\ }\bibfield  {title} {\enquote {\bibinfo {title} {{Enhanced prospects for direct detection of inelastic dark matter from a non-galactic diffuse component}},}\ }\href {\doibase 10.1088/1475-7516/2023/04/026} {\bibfield  {journal} {\bibinfo  {journal} {JCAP}\ }\textbf {\bibinfo {volume} {04}},\ \bibinfo {pages} {026} (\bibinfo {year} {2023})},\ \Eprint {http://arxiv.org/abs/2301.00870} {arXiv:2301.00870 [hep-ph]} \BibitemShut {NoStop}%
\bibitem [{\citenamefont {Gustafson}\ \emph {et~al.}(2025{\natexlab{b}})\citenamefont {Gustafson}, \citenamefont {Herrera}, \citenamefont {Mukhopadhyay}, \citenamefont {Murase},\ and\ \citenamefont {Shoemaker}}]{Gustafson:2024aom}%
  \BibitemOpen
  \bibfield  {author} {\bibinfo {author} {\bibfnamefont {R.~Andrew}\ \bibnamefont {Gustafson}}, \bibinfo {author} {\bibfnamefont {Gonzalo}\ \bibnamefont {Herrera}}, \bibinfo {author} {\bibfnamefont {Mainak}\ \bibnamefont {Mukhopadhyay}}, \bibinfo {author} {\bibfnamefont {Kohta}\ \bibnamefont {Murase}}, \ and\ \bibinfo {author} {\bibfnamefont {Ian~M.}\ \bibnamefont {Shoemaker}},\ }\bibfield  {title} {\enquote {\bibinfo {title} {{Cosmic-ray cooling in active galactic nuclei as a new probe of inelastic dark matter}},}\ }\href {\doibase 10.1103/5m57-vgt2} {\bibfield  {journal} {\bibinfo  {journal} {Phys. Rev. D}\ }\textbf {\bibinfo {volume} {111}},\ \bibinfo {pages} {L121303} (\bibinfo {year} {2025}{\natexlab{b}})},\ \Eprint {http://arxiv.org/abs/2408.08947} {arXiv:2408.08947 [hep-ph]} \BibitemShut {NoStop}%
\bibitem [{\citenamefont {Lee}\ \emph {et~al.}(2014)\citenamefont {Lee}, \citenamefont {Park},\ and\ \citenamefont {Sanz}}]{Lee:2013bua}%
  \BibitemOpen
  \bibfield  {author} {\bibinfo {author} {\bibfnamefont {Hyun~Min}\ \bibnamefont {Lee}}, \bibinfo {author} {\bibfnamefont {Myeonghun}\ \bibnamefont {Park}}, \ and\ \bibinfo {author} {\bibfnamefont {Veronica}\ \bibnamefont {Sanz}},\ }\bibfield  {title} {\enquote {\bibinfo {title} {{Gravity-mediated (or Composite) Dark Matter}},}\ }\href {\doibase 10.1140/epjc/s10052-014-2715-8} {\bibfield  {journal} {\bibinfo  {journal} {Eur. Phys. J. C}\ }\textbf {\bibinfo {volume} {74}},\ \bibinfo {pages} {2715} (\bibinfo {year} {2014})},\ \Eprint {http://arxiv.org/abs/1306.4107} {arXiv:1306.4107 [hep-ph]} \BibitemShut {NoStop}%
\bibitem [{\citenamefont {Kang}\ and\ \citenamefont {Lee}(2020)}]{Kang:2020huh}%
  \BibitemOpen
  \bibfield  {author} {\bibinfo {author} {\bibfnamefont {Yoo-Jin}\ \bibnamefont {Kang}}\ and\ \bibinfo {author} {\bibfnamefont {Hyun~Min}\ \bibnamefont {Lee}},\ }\bibfield  {title} {\enquote {\bibinfo {title} {{Lightening Gravity-Mediated Dark Matter}},}\ }\href {\doibase 10.1140/epjc/s10052-020-8153-x} {\bibfield  {journal} {\bibinfo  {journal} {Eur. Phys. J. C}\ }\textbf {\bibinfo {volume} {80}},\ \bibinfo {pages} {602} (\bibinfo {year} {2020})},\ \Eprint {http://arxiv.org/abs/2001.04868} {arXiv:2001.04868 [hep-ph]} \BibitemShut {NoStop}%
\bibitem [{\citenamefont {Gill}\ \emph {et~al.}(2023)\citenamefont {Gill}, \citenamefont {Sengupta},\ and\ \citenamefont {Williams}}]{Gill:2023kyz}%
  \BibitemOpen
  \bibfield  {author} {\bibinfo {author} {\bibfnamefont {Joshua~A.}\ \bibnamefont {Gill}}, \bibinfo {author} {\bibfnamefont {Dipan}\ \bibnamefont {Sengupta}}, \ and\ \bibinfo {author} {\bibfnamefont {Anthony~G.}\ \bibnamefont {Williams}},\ }\bibfield  {title} {\enquote {\bibinfo {title} {{Graviton-photon production with a massive spin-2 particle}},}\ }\href {\doibase 10.1103/PhysRevD.108.L051702} {\bibfield  {journal} {\bibinfo  {journal} {Phys. Rev. D}\ }\textbf {\bibinfo {volume} {108}},\ \bibinfo {pages} {L051702} (\bibinfo {year} {2023})},\ \Eprint {http://arxiv.org/abs/2303.04329} {arXiv:2303.04329 [hep-ph]} \BibitemShut {NoStop}%
\bibitem [{\citenamefont {Wang}\ \emph {et~al.}(2020)\citenamefont {Wang}, \citenamefont {Wu}, \citenamefont {Yang}, \citenamefont {Zhou},\ and\ \citenamefont {Zhu}}]{Wang:2019jtk}%
  \BibitemOpen
  \bibfield  {author} {\bibinfo {author} {\bibfnamefont {Wenyu}\ \bibnamefont {Wang}}, \bibinfo {author} {\bibfnamefont {Lei}\ \bibnamefont {Wu}}, \bibinfo {author} {\bibfnamefont {Jin~Min}\ \bibnamefont {Yang}}, \bibinfo {author} {\bibfnamefont {Hang}\ \bibnamefont {Zhou}}, \ and\ \bibinfo {author} {\bibfnamefont {Bin}\ \bibnamefont {Zhu}},\ }\bibfield  {title} {\enquote {\bibinfo {title} {{Cosmic ray boosted sub-GeV gravitationally interacting dark matter in direct detection}},}\ }\href {\doibase 10.1007/JHEP12(2020)072} {\bibfield  {journal} {\bibinfo  {journal} {JHEP}\ }\textbf {\bibinfo {volume} {12}},\ \bibinfo {pages} {072} (\bibinfo {year} {2020})},\ \bibinfo {note} {[Erratum: JHEP 02, 052 (2021)]},\ \Eprint {http://arxiv.org/abs/1912.09904} {arXiv:1912.09904 [hep-ph]} \BibitemShut {NoStop}%
\bibitem [{\citenamefont {de~Giorgi}\ and\ \citenamefont {Vogl}(2023)}]{deGiorgi:2022yha}%
  \BibitemOpen
  \bibfield  {author} {\bibinfo {author} {\bibfnamefont {Arturo}\ \bibnamefont {de~Giorgi}}\ and\ \bibinfo {author} {\bibfnamefont {Stefan}\ \bibnamefont {Vogl}},\ }\bibfield  {title} {\enquote {\bibinfo {title} {{Warm dark matter from a gravitational freeze-in in extra dimensions}},}\ }\href {\doibase 10.1007/JHEP04(2023)032} {\bibfield  {journal} {\bibinfo  {journal} {JHEP}\ }\textbf {\bibinfo {volume} {04}},\ \bibinfo {pages} {032} (\bibinfo {year} {2023})},\ \Eprint {http://arxiv.org/abs/2208.03153} {arXiv:2208.03153 [hep-ph]} \BibitemShut {NoStop}%
\bibitem [{\citenamefont {Voronchikhin}\ and\ \citenamefont {Kirpichnikov}(2025{\natexlab{b}})}]{Voronchikhin:2024ygo}%
  \BibitemOpen
  \bibfield  {author} {\bibinfo {author} {\bibfnamefont {I.~V.}\ \bibnamefont {Voronchikhin}}\ and\ \bibinfo {author} {\bibfnamefont {D.~V.}\ \bibnamefont {Kirpichnikov}},\ }\bibfield  {title} {\enquote {\bibinfo {title} {{The bremsstrahlung-like production of the massive spin-2 dark matter mediator}},}\ }\href {\doibase 10.1140/epjc/s10052-025-14868-6} {\bibfield  {journal} {\bibinfo  {journal} {Eur. Phys. J. C}\ }\textbf {\bibinfo {volume} {85}},\ \bibinfo {pages} {1110} (\bibinfo {year} {2025}{\natexlab{b}})},\ \Eprint {http://arxiv.org/abs/2412.10150} {arXiv:2412.10150 [hep-ph]} \BibitemShut {NoStop}%
\bibitem [{\citenamefont {Liu}\ and\ \citenamefont {Gurrola}(2026)}]{Liu:2026tsl}%
  \BibitemOpen
  \bibfield  {author} {\bibinfo {author} {\bibfnamefont {Junzhe}\ \bibnamefont {Liu}}\ and\ \bibinfo {author} {\bibfnamefont {Alfredo}\ \bibnamefont {Gurrola}},\ }\bibfield  {title} {\enquote {\bibinfo {title} {{Probing Freeze-In Dark Matter via a Spin-2 Portal at the LHC with Vector Boson Fusion and Machine Learning}},}\ }\href@noop {} {\  (\bibinfo {year} {2026})},\ \Eprint {http://arxiv.org/abs/2604.02604} {arXiv:2604.02604 [hep-ph]} \BibitemShut {NoStop}%
\bibitem [{\citenamefont {Tucker-Smith}\ and\ \citenamefont {Weiner}(2001)}]{Tucker-Smith:2001myb}%
  \BibitemOpen
  \bibfield  {author} {\bibinfo {author} {\bibfnamefont {David}\ \bibnamefont {Tucker-Smith}}\ and\ \bibinfo {author} {\bibfnamefont {Neal}\ \bibnamefont {Weiner}},\ }\bibfield  {title} {\enquote {\bibinfo {title} {{Inelastic dark matter}},}\ }\href {\doibase 10.1103/PhysRevD.64.043502} {\bibfield  {journal} {\bibinfo  {journal} {Phys. Rev. D}\ }\textbf {\bibinfo {volume} {64}},\ \bibinfo {pages} {043502} (\bibinfo {year} {2001})},\ \Eprint {http://arxiv.org/abs/hep-ph/0101138} {arXiv:hep-ph/0101138} \BibitemShut {NoStop}%
\bibitem [{\citenamefont {Bernabei}\ \emph {et~al.}(2013)\citenamefont {Bernabei} \emph {et~al.}}]{Bernabei:2013xsa}%
  \BibitemOpen
  \bibfield  {author} {\bibinfo {author} {\bibfnamefont {R.}~\bibnamefont {Bernabei}} \emph {et~al.},\ }\bibfield  {title} {\enquote {\bibinfo {title} {{Final model independent result of DAMA/LIBRA-phase1}},}\ }\href {\doibase 10.1140/epjc/s10052-013-2648-7} {\bibfield  {journal} {\bibinfo  {journal} {Eur. Phys. J. C}\ }\textbf {\bibinfo {volume} {73}},\ \bibinfo {pages} {2648} (\bibinfo {year} {2013})},\ \Eprint {http://arxiv.org/abs/1308.5109} {arXiv:1308.5109 [astro-ph.GA]} \BibitemShut {NoStop}%
\bibitem [{\citenamefont {De~Simone}\ \emph {et~al.}(2010)\citenamefont {De~Simone}, \citenamefont {Sanz},\ and\ \citenamefont {Sato}}]{DeSimone:2010tf}%
  \BibitemOpen
  \bibfield  {author} {\bibinfo {author} {\bibfnamefont {Andrea}\ \bibnamefont {De~Simone}}, \bibinfo {author} {\bibfnamefont {Veronica}\ \bibnamefont {Sanz}}, \ and\ \bibinfo {author} {\bibfnamefont {Hiromitsu~Phil}\ \bibnamefont {Sato}},\ }\bibfield  {title} {\enquote {\bibinfo {title} {{Pseudo-Dirac Dark Matter Leaves a Trace}},}\ }\href {\doibase 10.1103/PhysRevLett.105.121802} {\bibfield  {journal} {\bibinfo  {journal} {Phys. Rev. Lett.}\ }\textbf {\bibinfo {volume} {105}},\ \bibinfo {pages} {121802} (\bibinfo {year} {2010})},\ \Eprint {http://arxiv.org/abs/1004.1567} {arXiv:1004.1567 [hep-ph]} \BibitemShut {NoStop}%
\bibitem [{\citenamefont {Baryakhtar}\ \emph {et~al.}(2022)\citenamefont {Baryakhtar}, \citenamefont {Berlin}, \citenamefont {Liu},\ and\ \citenamefont {Weiner}}]{Baryakhtar:2020rwy}%
  \BibitemOpen
  \bibfield  {author} {\bibinfo {author} {\bibfnamefont {Masha}\ \bibnamefont {Baryakhtar}}, \bibinfo {author} {\bibfnamefont {Asher}\ \bibnamefont {Berlin}}, \bibinfo {author} {\bibfnamefont {Hongwan}\ \bibnamefont {Liu}}, \ and\ \bibinfo {author} {\bibfnamefont {Neal}\ \bibnamefont {Weiner}},\ }\bibfield  {title} {\enquote {\bibinfo {title} {{Electromagnetic signals of inelastic dark matter scattering}},}\ }\href {\doibase 10.1007/JHEP06(2022)047} {\bibfield  {journal} {\bibinfo  {journal} {JHEP}\ }\textbf {\bibinfo {volume} {06}},\ \bibinfo {pages} {047} (\bibinfo {year} {2022})},\ \Eprint {http://arxiv.org/abs/2006.13918} {arXiv:2006.13918 [hep-ph]} \BibitemShut {NoStop}%
\bibitem [{\citenamefont {Carrillo~Gonz{\'a}lez}\ and\ \citenamefont {Toro}(2022)}]{CarrilloGonzalez:2021lxm}%
  \BibitemOpen
  \bibfield  {author} {\bibinfo {author} {\bibfnamefont {Mariana}\ \bibnamefont {Carrillo~Gonz{\'a}lez}}\ and\ \bibinfo {author} {\bibfnamefont {Natalia}\ \bibnamefont {Toro}},\ }\bibfield  {title} {\enquote {\bibinfo {title} {{Cosmology and signals of light pseudo-Dirac dark matter}},}\ }\href {\doibase 10.1007/JHEP04(2022)060} {\bibfield  {journal} {\bibinfo  {journal} {JHEP}\ }\textbf {\bibinfo {volume} {04}},\ \bibinfo {pages} {060} (\bibinfo {year} {2022})},\ \Eprint {http://arxiv.org/abs/2108.13422} {arXiv:2108.13422 [hep-ph]} \BibitemShut {NoStop}%
\bibitem [{\citenamefont {Izaguirre}\ \emph {et~al.}(2017)\citenamefont {Izaguirre}, \citenamefont {Kahn}, \citenamefont {Krnjaic},\ and\ \citenamefont {Moschella}}]{Izaguirre:2017bqb}%
  \BibitemOpen
  \bibfield  {author} {\bibinfo {author} {\bibfnamefont {Eder}\ \bibnamefont {Izaguirre}}, \bibinfo {author} {\bibfnamefont {Yonatan}\ \bibnamefont {Kahn}}, \bibinfo {author} {\bibfnamefont {Gordan}\ \bibnamefont {Krnjaic}}, \ and\ \bibinfo {author} {\bibfnamefont {Matthew}\ \bibnamefont {Moschella}},\ }\bibfield  {title} {\enquote {\bibinfo {title} {{Testing Light Dark Matter Coannihilation With Fixed-Target Experiments}},}\ }\href {\doibase 10.1103/PhysRevD.96.055007} {\bibfield  {journal} {\bibinfo  {journal} {Phys. Rev. D}\ }\textbf {\bibinfo {volume} {96}},\ \bibinfo {pages} {055007} (\bibinfo {year} {2017})},\ \Eprint {http://arxiv.org/abs/1703.06881} {arXiv:1703.06881 [hep-ph]} \BibitemShut {NoStop}%
\bibitem [{\citenamefont {Foguel}\ \emph {et~al.}(2025)\citenamefont {Foguel}, \citenamefont {Reimitz},\ and\ \citenamefont {Funchal}}]{Foguel:2024lca}%
  \BibitemOpen
  \bibfield  {author} {\bibinfo {author} {\bibfnamefont {Ana~Luisa}\ \bibnamefont {Foguel}}, \bibinfo {author} {\bibfnamefont {Peter}\ \bibnamefont {Reimitz}}, \ and\ \bibinfo {author} {\bibfnamefont {Renata~Zukanovich}\ \bibnamefont {Funchal}},\ }\bibfield  {title} {\enquote {\bibinfo {title} {{Unlocking the inelastic Dark Matter window with vector mediators}},}\ }\href {\doibase 10.1007/JHEP05(2025)001} {\bibfield  {journal} {\bibinfo  {journal} {JHEP}\ }\textbf {\bibinfo {volume} {05}},\ \bibinfo {pages} {001} (\bibinfo {year} {2025})},\ \Eprint {http://arxiv.org/abs/2410.00881} {arXiv:2410.00881 [hep-ph]} \BibitemShut {NoStop}%
\bibitem [{\citenamefont {Voronchikhin}\ and\ \citenamefont {Kirpichnikov}(2026)}]{Voronchikhin:2025eqm}%
  \BibitemOpen
  \bibfield  {author} {\bibinfo {author} {\bibfnamefont {I.~V.}\ \bibnamefont {Voronchikhin}}\ and\ \bibinfo {author} {\bibfnamefont {D.~V.}\ \bibnamefont {Kirpichnikov}},\ }\bibfield  {title} {\enquote {\bibinfo {title} {{Examining scalar portal inelastic dark matter with lepton fixed-target experiments}},}\ }\href {\doibase 10.1103/qvbp-rhsr} {\bibfield  {journal} {\bibinfo  {journal} {Phys. Rev. D}\ }\textbf {\bibinfo {volume} {113}},\ \bibinfo {pages} {015031} (\bibinfo {year} {2026})},\ \Eprint {http://arxiv.org/abs/2505.04290} {arXiv:2505.04290 [hep-ph]} \BibitemShut {NoStop}%
\bibitem [{\citenamefont {Gninenko}\ \emph {et~al.}(2026)\citenamefont {Gninenko}, \citenamefont {Krasnikov}, \citenamefont {Voronchikhin},\ and\ \citenamefont {Kirpichnikov}}]{Gninenko:2026svu}%
  \BibitemOpen
  \bibfield  {author} {\bibinfo {author} {\bibfnamefont {Sergei~N.}\ \bibnamefont {Gninenko}}, \bibinfo {author} {\bibfnamefont {N.~V.}\ \bibnamefont {Krasnikov}}, \bibinfo {author} {\bibfnamefont {I.~V.}\ \bibnamefont {Voronchikhin}}, \ and\ \bibinfo {author} {\bibfnamefont {D.~V.}\ \bibnamefont {Kirpichnikov}},\ }\bibfield  {title} {\enquote {\bibinfo {title} {{Missing energy signatures of inelastic magnetic dipole DM at NA64e}},}\ }\href@noop {} {\  (\bibinfo {year} {2026})},\ \Eprint {http://arxiv.org/abs/2603.28278} {arXiv:2603.28278 [hep-ph]} \BibitemShut {NoStop}%
\bibitem [{\citenamefont {Berlin}\ and\ \citenamefont {Kling}(2019)}]{Berlin:2018jbm}%
  \BibitemOpen
  \bibfield  {author} {\bibinfo {author} {\bibfnamefont {Asher}\ \bibnamefont {Berlin}}\ and\ \bibinfo {author} {\bibfnamefont {Felix}\ \bibnamefont {Kling}},\ }\bibfield  {title} {\enquote {\bibinfo {title} {{Inelastic Dark Matter at the LHC Lifetime Frontier: ATLAS, CMS, LHCb, CODEX-b, FASER, and MATHUSLA}},}\ }\href {\doibase 10.1103/PhysRevD.99.015021} {\bibfield  {journal} {\bibinfo  {journal} {Phys. Rev. D}\ }\textbf {\bibinfo {volume} {99}},\ \bibinfo {pages} {015021} (\bibinfo {year} {2019})},\ \Eprint {http://arxiv.org/abs/1810.01879} {arXiv:1810.01879 [hep-ph]} \BibitemShut {NoStop}%
\bibitem [{\citenamefont {Jod{\l}owski}(2023)}]{Jodlowski:2023ohn}%
  \BibitemOpen
  \bibfield  {author} {\bibinfo {author} {\bibfnamefont {Krzysztof}\ \bibnamefont {Jod{\l}owski}},\ }\bibfield  {title} {\enquote {\bibinfo {title} {{Looking forward to inelastic DM with electromagnetic form factors at FASER and beam dump experiments}},}\ }\href {\doibase 10.1103/PhysRevD.108.115025} {\bibfield  {journal} {\bibinfo  {journal} {Phys. Rev. D}\ }\textbf {\bibinfo {volume} {108}},\ \bibinfo {pages} {115025} (\bibinfo {year} {2023})},\ \Eprint {http://arxiv.org/abs/2305.16781} {arXiv:2305.16781 [hep-ph]} \BibitemShut {NoStop}%
\bibitem [{\citenamefont {Dienes}\ \emph {et~al.}(2023)\citenamefont {Dienes}, \citenamefont {Feng}, \citenamefont {Fieg}, \citenamefont {Huang}, \citenamefont {Lee},\ and\ \citenamefont {Thomas}}]{Dienes:2023uve}%
  \BibitemOpen
  \bibfield  {author} {\bibinfo {author} {\bibfnamefont {Keith~R.}\ \bibnamefont {Dienes}}, \bibinfo {author} {\bibfnamefont {Jonathan~L.}\ \bibnamefont {Feng}}, \bibinfo {author} {\bibfnamefont {Max}\ \bibnamefont {Fieg}}, \bibinfo {author} {\bibfnamefont {Fei}\ \bibnamefont {Huang}}, \bibinfo {author} {\bibfnamefont {Seung~J.}\ \bibnamefont {Lee}}, \ and\ \bibinfo {author} {\bibfnamefont {Brooks}\ \bibnamefont {Thomas}},\ }\bibfield  {title} {\enquote {\bibinfo {title} {{Extending the discovery potential for inelastic-dipole dark matter with FASER}},}\ }\href {\doibase 10.1103/PhysRevD.107.115006} {\bibfield  {journal} {\bibinfo  {journal} {Phys. Rev. D}\ }\textbf {\bibinfo {volume} {107}},\ \bibinfo {pages} {115006} (\bibinfo {year} {2023})},\ \Eprint {http://arxiv.org/abs/2301.05252} {arXiv:2301.05252 [hep-ph]} \BibitemShut {NoStop}%
\bibitem [{\citenamefont {Goodman}\ and\ \citenamefont {Witten}(1985)}]{Goodman:1984dc}%
  \BibitemOpen
  \bibfield  {author} {\bibinfo {author} {\bibfnamefont {Mark~W.}\ \bibnamefont {Goodman}}\ and\ \bibinfo {author} {\bibfnamefont {Edward}\ \bibnamefont {Witten}},\ }\bibfield  {title} {\enquote {\bibinfo {title} {{Detectability of Certain Dark Matter Candidates}},}\ }\href {\doibase 10.1103/PhysRevD.31.3059} {\bibfield  {journal} {\bibinfo  {journal} {Phys. Rev. D}\ }\textbf {\bibinfo {volume} {31}},\ \bibinfo {pages} {3059} (\bibinfo {year} {1985})}\BibitemShut {NoStop}%
\bibitem [{\citenamefont {Schumann}(2019)}]{Schumann:2019eaa}%
  \BibitemOpen
  \bibfield  {author} {\bibinfo {author} {\bibfnamefont {Marc}\ \bibnamefont {Schumann}},\ }\bibfield  {title} {\enquote {\bibinfo {title} {{Direct Detection of WIMP Dark Matter: Concepts and Status}},}\ }\href {\doibase 10.1088/1361-6471/ab2ea5} {\bibfield  {journal} {\bibinfo  {journal} {J. Phys. G}\ }\textbf {\bibinfo {volume} {46}},\ \bibinfo {pages} {103003} (\bibinfo {year} {2019})},\ \Eprint {http://arxiv.org/abs/1903.03026} {arXiv:1903.03026 [astro-ph.CO]} \BibitemShut {NoStop}%
\bibitem [{\citenamefont {Essig}\ \emph {et~al.}(2012{\natexlab{a}})\citenamefont {Essig}, \citenamefont {Mardon},\ and\ \citenamefont {Volansky}}]{Essig:2011nj}%
  \BibitemOpen
  \bibfield  {author} {\bibinfo {author} {\bibfnamefont {Rouven}\ \bibnamefont {Essig}}, \bibinfo {author} {\bibfnamefont {Jeremy}\ \bibnamefont {Mardon}}, \ and\ \bibinfo {author} {\bibfnamefont {Tomer}\ \bibnamefont {Volansky}},\ }\bibfield  {title} {\enquote {\bibinfo {title} {{Direct Detection of Sub-GeV Dark Matter}},}\ }\href {\doibase 10.1103/PhysRevD.85.076007} {\bibfield  {journal} {\bibinfo  {journal} {Phys. Rev. D}\ }\textbf {\bibinfo {volume} {85}},\ \bibinfo {pages} {076007} (\bibinfo {year} {2012}{\natexlab{a}})},\ \Eprint {http://arxiv.org/abs/1108.5383} {arXiv:1108.5383 [hep-ph]} \BibitemShut {NoStop}%
\bibitem [{\citenamefont {Essig}\ \emph {et~al.}(2012{\natexlab{b}})\citenamefont {Essig}, \citenamefont {Manalaysay}, \citenamefont {Mardon}, \citenamefont {Sorensen},\ and\ \citenamefont {Volansky}}]{Essig:2012yx}%
  \BibitemOpen
  \bibfield  {author} {\bibinfo {author} {\bibfnamefont {Rouven}\ \bibnamefont {Essig}}, \bibinfo {author} {\bibfnamefont {Aaron}\ \bibnamefont {Manalaysay}}, \bibinfo {author} {\bibfnamefont {Jeremy}\ \bibnamefont {Mardon}}, \bibinfo {author} {\bibfnamefont {Peter}\ \bibnamefont {Sorensen}}, \ and\ \bibinfo {author} {\bibfnamefont {Tomer}\ \bibnamefont {Volansky}},\ }\bibfield  {title} {\enquote {\bibinfo {title} {{First Direct Detection Limits on sub-GeV Dark Matter from XENON10}},}\ }\href {\doibase 10.1103/PhysRevLett.109.021301} {\bibfield  {journal} {\bibinfo  {journal} {Phys. Rev. Lett.}\ }\textbf {\bibinfo {volume} {109}},\ \bibinfo {pages} {021301} (\bibinfo {year} {2012}{\natexlab{b}})},\ \Eprint {http://arxiv.org/abs/1206.2644} {arXiv:1206.2644 [astro-ph.CO]} \BibitemShut {NoStop}%
\bibitem [{\citenamefont {Essig}\ \emph {et~al.}(2017)\citenamefont {Essig}, \citenamefont {Volansky},\ and\ \citenamefont {Yu}}]{Essig:2017kqs}%
  \BibitemOpen
  \bibfield  {author} {\bibinfo {author} {\bibfnamefont {Rouven}\ \bibnamefont {Essig}}, \bibinfo {author} {\bibfnamefont {Tomer}\ \bibnamefont {Volansky}}, \ and\ \bibinfo {author} {\bibfnamefont {Tien-Tien}\ \bibnamefont {Yu}},\ }\bibfield  {title} {\enquote {\bibinfo {title} {{New Constraints and Prospects for sub-GeV Dark Matter Scattering off Electrons in Xenon}},}\ }\href {\doibase 10.1103/PhysRevD.96.043017} {\bibfield  {journal} {\bibinfo  {journal} {Phys. Rev. D}\ }\textbf {\bibinfo {volume} {96}},\ \bibinfo {pages} {043017} (\bibinfo {year} {2017})},\ \Eprint {http://arxiv.org/abs/1703.00910} {arXiv:1703.00910 [hep-ph]} \BibitemShut {NoStop}%
\bibitem [{\citenamefont {Emken}\ \emph {et~al.}(2019)\citenamefont {Emken}, \citenamefont {Essig}, \citenamefont {Kouvaris},\ and\ \citenamefont {Sholapurkar}}]{Emken:2019tni}%
  \BibitemOpen
  \bibfield  {author} {\bibinfo {author} {\bibfnamefont {Timon}\ \bibnamefont {Emken}}, \bibinfo {author} {\bibfnamefont {Rouven}\ \bibnamefont {Essig}}, \bibinfo {author} {\bibfnamefont {Chris}\ \bibnamefont {Kouvaris}}, \ and\ \bibinfo {author} {\bibfnamefont {Mukul}\ \bibnamefont {Sholapurkar}},\ }\bibfield  {title} {\enquote {\bibinfo {title} {{Direct Detection of Strongly Interacting Sub-GeV Dark Matter via Electron Recoils}},}\ }\href {\doibase 10.1088/1475-7516/2019/09/070} {\bibfield  {journal} {\bibinfo  {journal} {JCAP}\ }\textbf {\bibinfo {volume} {09}},\ \bibinfo {pages} {070} (\bibinfo {year} {2019})},\ \Eprint {http://arxiv.org/abs/1905.06348} {arXiv:1905.06348 [hep-ph]} \BibitemShut {NoStop}%
\bibitem [{\citenamefont {Essig}\ \emph {et~al.}(2016)\citenamefont {Essig}, \citenamefont {Fernandez-Serra}, \citenamefont {Mardon}, \citenamefont {Soto}, \citenamefont {Volansky},\ and\ \citenamefont {Yu}}]{Essig:2015cda}%
  \BibitemOpen
  \bibfield  {author} {\bibinfo {author} {\bibfnamefont {Rouven}\ \bibnamefont {Essig}}, \bibinfo {author} {\bibfnamefont {Marivi}\ \bibnamefont {Fernandez-Serra}}, \bibinfo {author} {\bibfnamefont {Jeremy}\ \bibnamefont {Mardon}}, \bibinfo {author} {\bibfnamefont {Adrian}\ \bibnamefont {Soto}}, \bibinfo {author} {\bibfnamefont {Tomer}\ \bibnamefont {Volansky}}, \ and\ \bibinfo {author} {\bibfnamefont {Tien-Tien}\ \bibnamefont {Yu}},\ }\bibfield  {title} {\enquote {\bibinfo {title} {{Direct Detection of sub-GeV Dark Matter with Semiconductor Targets}},}\ }\href {\doibase 10.1007/JHEP05(2016)046} {\bibfield  {journal} {\bibinfo  {journal} {JHEP}\ }\textbf {\bibinfo {volume} {05}},\ \bibinfo {pages} {046} (\bibinfo {year} {2016})},\ \Eprint {http://arxiv.org/abs/1509.01598} {arXiv:1509.01598 [hep-ph]} \BibitemShut {NoStop}%
\bibitem [{\citenamefont {Caddell}\ \emph {et~al.}(2023)\citenamefont {Caddell}, \citenamefont {Flambaum},\ and\ \citenamefont {Roberts}}]{Caddell:2023zsw}%
  \BibitemOpen
  \bibfield  {author} {\bibinfo {author} {\bibfnamefont {A.~R.}\ \bibnamefont {Caddell}}, \bibinfo {author} {\bibfnamefont {V.~V.}\ \bibnamefont {Flambaum}}, \ and\ \bibinfo {author} {\bibfnamefont {B.~M.}\ \bibnamefont {Roberts}},\ }\bibfield  {title} {\enquote {\bibinfo {title} {{Accurate electron-recoil ionization factors for dark matter direct detection in xenon, krypton, and argon}},}\ }\href {\doibase 10.1103/PhysRevD.108.083030} {\bibfield  {journal} {\bibinfo  {journal} {Phys. Rev. D}\ }\textbf {\bibinfo {volume} {108}},\ \bibinfo {pages} {083030} (\bibinfo {year} {2023})},\ \Eprint {http://arxiv.org/abs/2305.05125} {arXiv:2305.05125 [hep-ph]} \BibitemShut {NoStop}%
\bibitem [{\citenamefont {Finkbeiner}\ \emph {et~al.}(2009)\citenamefont {Finkbeiner}, \citenamefont {Slatyer}, \citenamefont {Weiner},\ and\ \citenamefont {Yavin}}]{Finkbeiner:2009mi}%
  \BibitemOpen
  \bibfield  {author} {\bibinfo {author} {\bibfnamefont {Douglas~P.}\ \bibnamefont {Finkbeiner}}, \bibinfo {author} {\bibfnamefont {Tracy~R.}\ \bibnamefont {Slatyer}}, \bibinfo {author} {\bibfnamefont {Neal}\ \bibnamefont {Weiner}}, \ and\ \bibinfo {author} {\bibfnamefont {Itay}\ \bibnamefont {Yavin}},\ }\bibfield  {title} {\enquote {\bibinfo {title} {{PAMELA, DAMA, INTEGRAL and Signatures of Metastable Excited WIMPs}},}\ }\href {\doibase 10.1088/1475-7516/2009/09/037} {\bibfield  {journal} {\bibinfo  {journal} {JCAP}\ }\textbf {\bibinfo {volume} {09}},\ \bibinfo {pages} {037} (\bibinfo {year} {2009})},\ \Eprint {http://arxiv.org/abs/0903.1037} {arXiv:0903.1037 [hep-ph]} \BibitemShut {NoStop}%
\bibitem [{\citenamefont {Harigaya}\ \emph {et~al.}(2020)\citenamefont {Harigaya}, \citenamefont {Nakai},\ and\ \citenamefont {Suzuki}}]{Harigaya:2020ckz}%
  \BibitemOpen
  \bibfield  {author} {\bibinfo {author} {\bibfnamefont {Keisuke}\ \bibnamefont {Harigaya}}, \bibinfo {author} {\bibfnamefont {Yuichiro}\ \bibnamefont {Nakai}}, \ and\ \bibinfo {author} {\bibfnamefont {Motoo}\ \bibnamefont {Suzuki}},\ }\bibfield  {title} {\enquote {\bibinfo {title} {{Inelastic Dark Matter Electron Scattering and the XENON1T Excess}},}\ }\href {\doibase 10.1016/j.physletb.2020.135729} {\bibfield  {journal} {\bibinfo  {journal} {Phys. Lett. B}\ }\textbf {\bibinfo {volume} {809}},\ \bibinfo {pages} {135729} (\bibinfo {year} {2020})},\ \Eprint {http://arxiv.org/abs/2006.11938} {arXiv:2006.11938 [hep-ph]} \BibitemShut {NoStop}%
\bibitem [{\citenamefont {Aprile}\ \emph {et~al.}(2020)\citenamefont {Aprile} \emph {et~al.}}]{XENON:2020rca}%
  \BibitemOpen
  \bibfield  {author} {\bibinfo {author} {\bibfnamefont {E.}~\bibnamefont {Aprile}} \emph {et~al.} (\bibinfo {collaboration} {XENON}),\ }\bibfield  {title} {\enquote {\bibinfo {title} {{Excess electronic recoil events in XENON1T}},}\ }\href {\doibase 10.1103/PhysRevD.102.072004} {\bibfield  {journal} {\bibinfo  {journal} {Phys. Rev. D}\ }\textbf {\bibinfo {volume} {102}},\ \bibinfo {pages} {072004} (\bibinfo {year} {2020})},\ \Eprint {http://arxiv.org/abs/2006.09721} {arXiv:2006.09721 [hep-ex]} \BibitemShut {NoStop}%
\bibitem [{\citenamefont {He}\ \emph {et~al.}(2024)\citenamefont {He}, \citenamefont {Wang},\ and\ \citenamefont {Zheng}}]{He:2024hkr}%
  \BibitemOpen
  \bibfield  {author} {\bibinfo {author} {\bibfnamefont {Hong-Jian}\ \bibnamefont {He}}, \bibinfo {author} {\bibfnamefont {Yu-Chen}\ \bibnamefont {Wang}}, \ and\ \bibinfo {author} {\bibfnamefont {Jiaming}\ \bibnamefont {Zheng}},\ }\bibfield  {title} {\enquote {\bibinfo {title} {{Probing light inelastic dark matter from direct detection}},}\ }\href {\doibase 10.1016/j.dark.2024.101670} {\bibfield  {journal} {\bibinfo  {journal} {Phys. Dark Univ.}\ }\textbf {\bibinfo {volume} {46}},\ \bibinfo {pages} {101670} (\bibinfo {year} {2024})},\ \Eprint {http://arxiv.org/abs/2403.03128} {arXiv:2403.03128 [hep-ph]} \BibitemShut {NoStop}%
\bibitem [{\citenamefont {Wang}\ \emph {et~al.}(2025)\citenamefont {Wang}, \citenamefont {Yun}, \citenamefont {He},\ and\ \citenamefont {Meng}}]{Wang:2025uwh}%
  \BibitemOpen
  \bibfield  {author} {\bibinfo {author} {\bibfnamefont {Yu-Chen}\ \bibnamefont {Wang}}, \bibinfo {author} {\bibfnamefont {Youhui}\ \bibnamefont {Yun}}, \bibinfo {author} {\bibfnamefont {Hong-Jian}\ \bibnamefont {He}}, \ and\ \bibinfo {author} {\bibfnamefont {Yue}\ \bibnamefont {Meng}},\ }\bibfield  {title} {\enquote {\bibinfo {title} {{Search for Light Inelastic Dark Matter with Low-Energy Ionization Signatures in PandaX-4T}},}\ }\href@noop {} {\  (\bibinfo {year} {2025})},\ \Eprint {http://arxiv.org/abs/2508.13062} {arXiv:2508.13062 [hep-ph]} \BibitemShut {NoStop}%
\bibitem [{\citenamefont {Brahma}\ \emph {et~al.}(2024)\citenamefont {Brahma}, \citenamefont {Heeba},\ and\ \citenamefont {Schutz}}]{Brahma:2023psr}%
  \BibitemOpen
  \bibfield  {author} {\bibinfo {author} {\bibfnamefont {Nirmalya}\ \bibnamefont {Brahma}}, \bibinfo {author} {\bibfnamefont {Saniya}\ \bibnamefont {Heeba}}, \ and\ \bibinfo {author} {\bibfnamefont {Katelin}\ \bibnamefont {Schutz}},\ }\bibfield  {title} {\enquote {\bibinfo {title} {{Resonant pseudo-Dirac dark matter as a sub-GeV thermal target}},}\ }\href {\doibase 10.1103/PhysRevD.109.035006} {\bibfield  {journal} {\bibinfo  {journal} {Phys. Rev. D}\ }\textbf {\bibinfo {volume} {109}},\ \bibinfo {pages} {035006} (\bibinfo {year} {2024})},\ \Eprint {http://arxiv.org/abs/2308.01960} {arXiv:2308.01960 [hep-ph]} \BibitemShut {NoStop}%
\bibitem [{\citenamefont {Berlin}\ \emph {et~al.}(2019)\citenamefont {Berlin}, \citenamefont {Blinov}, \citenamefont {Krnjaic}, \citenamefont {Schuster},\ and\ \citenamefont {Toro}}]{Berlin:2018bsc}%
  \BibitemOpen
  \bibfield  {author} {\bibinfo {author} {\bibfnamefont {Asher}\ \bibnamefont {Berlin}}, \bibinfo {author} {\bibfnamefont {Nikita}\ \bibnamefont {Blinov}}, \bibinfo {author} {\bibfnamefont {Gordan}\ \bibnamefont {Krnjaic}}, \bibinfo {author} {\bibfnamefont {Philip}\ \bibnamefont {Schuster}}, \ and\ \bibinfo {author} {\bibfnamefont {Natalia}\ \bibnamefont {Toro}},\ }\bibfield  {title} {\enquote {\bibinfo {title} {{Dark Matter, Millicharges, Axion and Scalar Particles, Gauge Bosons, and Other New Physics with LDMX}},}\ }\href {\doibase 10.1103/PhysRevD.99.075001} {\bibfield  {journal} {\bibinfo  {journal} {Phys. Rev. D}\ }\textbf {\bibinfo {volume} {99}},\ \bibinfo {pages} {075001} (\bibinfo {year} {2019})},\ \Eprint {http://arxiv.org/abs/1807.01730} {arXiv:1807.01730 [hep-ph]} \BibitemShut {NoStop}%
\bibitem [{\citenamefont {Chen}\ \emph {et~al.}(2018)\citenamefont {Chen}, \citenamefont {Kozaczuk},\ and\ \citenamefont {Zhong}}]{Chen:2018vkr}%
  \BibitemOpen
  \bibfield  {author} {\bibinfo {author} {\bibfnamefont {Chien-Yi}\ \bibnamefont {Chen}}, \bibinfo {author} {\bibfnamefont {Jonathan}\ \bibnamefont {Kozaczuk}}, \ and\ \bibinfo {author} {\bibfnamefont {Yi-Ming}\ \bibnamefont {Zhong}},\ }\bibfield  {title} {\enquote {\bibinfo {title} {{Exploring leptophilic dark matter with NA64-$\mu$}},}\ }\href {\doibase 10.1007/JHEP10(2018)154} {\bibfield  {journal} {\bibinfo  {journal} {JHEP}\ }\textbf {\bibinfo {volume} {10}},\ \bibinfo {pages} {154} (\bibinfo {year} {2018})},\ \Eprint {http://arxiv.org/abs/1807.03790} {arXiv:1807.03790 [hep-ph]} \BibitemShut {NoStop}%
\bibitem [{\citenamefont {Batell}\ \emph {et~al.}(2018)\citenamefont {Batell}, \citenamefont {Freitas}, \citenamefont {Ismail},\ and\ \citenamefont {Mckeen}}]{Batell:2017kty}%
  \BibitemOpen
  \bibfield  {author} {\bibinfo {author} {\bibfnamefont {Brian}\ \bibnamefont {Batell}}, \bibinfo {author} {\bibfnamefont {Ayres}\ \bibnamefont {Freitas}}, \bibinfo {author} {\bibfnamefont {Ahmed}\ \bibnamefont {Ismail}}, \ and\ \bibinfo {author} {\bibfnamefont {David}\ \bibnamefont {Mckeen}},\ }\bibfield  {title} {\enquote {\bibinfo {title} {{Flavor-specific scalar mediators}},}\ }\href {\doibase 10.1103/PhysRevD.98.055026} {\bibfield  {journal} {\bibinfo  {journal} {Phys. Rev. D}\ }\textbf {\bibinfo {volume} {98}},\ \bibinfo {pages} {055026} (\bibinfo {year} {2018})},\ \Eprint {http://arxiv.org/abs/1712.10022} {arXiv:1712.10022 [hep-ph]} \BibitemShut {NoStop}%
\bibitem [{\citenamefont {Dreiner}\ \emph {et~al.}(2010)\citenamefont {Dreiner}, \citenamefont {Haber},\ and\ \citenamefont {Martin}}]{Dreiner:2008tw}%
  \BibitemOpen
  \bibfield  {author} {\bibinfo {author} {\bibfnamefont {Herbi~K.}\ \bibnamefont {Dreiner}}, \bibinfo {author} {\bibfnamefont {Howard~E.}\ \bibnamefont {Haber}}, \ and\ \bibinfo {author} {\bibfnamefont {Stephen~P.}\ \bibnamefont {Martin}},\ }\bibfield  {title} {\enquote {\bibinfo {title} {{Two-component spinor techniques and Feynman rules for quantum field theory and supersymmetry}},}\ }\href {\doibase 10.1016/j.physrep.2010.05.002} {\bibfield  {journal} {\bibinfo  {journal} {Phys. Rept.}\ }\textbf {\bibinfo {volume} {494}},\ \bibinfo {pages} {1--196} (\bibinfo {year} {2010})},\ \Eprint {http://arxiv.org/abs/0812.1594} {arXiv:0812.1594 [hep-ph]} \BibitemShut {NoStop}%
\bibitem [{\citenamefont {Krnjaic}\ \emph {et~al.}(2025{\natexlab{a}})\citenamefont {Krnjaic}, \citenamefont {McKeen}, \citenamefont {Mizuta}, \citenamefont {Mohlabeng}, \citenamefont {Morrissey},\ and\ \citenamefont {Tuckler}}]{Krnjaic:2025zjl}%
  \BibitemOpen
  \bibfield  {author} {\bibinfo {author} {\bibfnamefont {Gordan}\ \bibnamefont {Krnjaic}}, \bibinfo {author} {\bibfnamefont {David}\ \bibnamefont {McKeen}}, \bibinfo {author} {\bibfnamefont {Riku}\ \bibnamefont {Mizuta}}, \bibinfo {author} {\bibfnamefont {Gopolang}\ \bibnamefont {Mohlabeng}}, \bibinfo {author} {\bibfnamefont {David~E.}\ \bibnamefont {Morrissey}}, \ and\ \bibinfo {author} {\bibfnamefont {Douglas}\ \bibnamefont {Tuckler}},\ }\bibfield  {title} {\enquote {\bibinfo {title} {{X-rays from inelastic dark matter freeze-in}},}\ }\href {\doibase 10.1103/99z7-kz4s} {\bibfield  {journal} {\bibinfo  {journal} {Phys. Rev. D}\ }\textbf {\bibinfo {volume} {112}},\ \bibinfo {pages} {115039} (\bibinfo {year} {2025}{\natexlab{a}})},\ \Eprint {http://arxiv.org/abs/2509.19428} {arXiv:2509.19428 [hep-ph]} \BibitemShut {NoStop}%
\bibitem [{\citenamefont {Dalla Valle~Garcia}(2025)}]{DallaValleGarcia:2024zva}%
  \BibitemOpen
  \bibfield  {author} {\bibinfo {author} {\bibfnamefont {Giovani}\ \bibnamefont {Dalla Valle~Garcia}},\ }\bibfield  {title} {\enquote {\bibinfo {title} {{A minimalistic model for inelastic dark matter}},}\ }\href {\doibase 10.1016/j.physletb.2025.139320} {\bibfield  {journal} {\bibinfo  {journal} {Phys. Lett. B}\ }\textbf {\bibinfo {volume} {862}},\ \bibinfo {pages} {139320} (\bibinfo {year} {2025})},\ \Eprint {http://arxiv.org/abs/2411.02147} {arXiv:2411.02147 [hep-ph]} \BibitemShut {NoStop}%
\bibitem [{\citenamefont {Griest}\ and\ \citenamefont {Seckel}(1991)}]{Griest:1990kh}%
  \BibitemOpen
  \bibfield  {author} {\bibinfo {author} {\bibfnamefont {Kim}\ \bibnamefont {Griest}}\ and\ \bibinfo {author} {\bibfnamefont {David}\ \bibnamefont {Seckel}},\ }\bibfield  {title} {\enquote {\bibinfo {title} {{Three exceptions in the calculation of relic abundances}},}\ }\href {\doibase 10.1103/PhysRevD.43.3191} {\bibfield  {journal} {\bibinfo  {journal} {Phys. Rev. D}\ }\textbf {\bibinfo {volume} {43}},\ \bibinfo {pages} {3191--3203} (\bibinfo {year} {1991})}\BibitemShut {NoStop}%
\bibitem [{\citenamefont {Edsjo}\ and\ \citenamefont {Gondolo}(1997)}]{Edsjo:1997bg}%
  \BibitemOpen
  \bibfield  {author} {\bibinfo {author} {\bibfnamefont {Joakim}\ \bibnamefont {Edsjo}}\ and\ \bibinfo {author} {\bibfnamefont {Paolo}\ \bibnamefont {Gondolo}},\ }\bibfield  {title} {\enquote {\bibinfo {title} {{Neutralino relic density including coannihilations}},}\ }\href {\doibase 10.1103/PhysRevD.56.1879} {\bibfield  {journal} {\bibinfo  {journal} {Phys. Rev. D}\ }\textbf {\bibinfo {volume} {56}},\ \bibinfo {pages} {1879--1894} (\bibinfo {year} {1997})},\ \Eprint {http://arxiv.org/abs/hep-ph/9704361} {arXiv:hep-ph/9704361} \BibitemShut {NoStop}%
\bibitem [{\citenamefont {Krnjaic}(2025)}]{Krnjaic:2025noj}%
  \BibitemOpen
  \bibfield  {author} {\bibinfo {author} {\bibfnamefont {Gordan}\ \bibnamefont {Krnjaic}},\ }\bibfield  {title} {\enquote {\bibinfo {title} {{Testing Thermal-Relic Dark Matter with a Dark Photon Mediator}},}\ }\href@noop {} {\  (\bibinfo {year} {2025})},\ \Eprint {http://arxiv.org/abs/2505.04626} {arXiv:2505.04626 [hep-ph]} \BibitemShut {NoStop}%
\bibitem [{\citenamefont {Husdal}(2016)}]{Husdal:2016haj}%
  \BibitemOpen
  \bibfield  {author} {\bibinfo {author} {\bibfnamefont {Lars}\ \bibnamefont {Husdal}},\ }\bibfield  {title} {\enquote {\bibinfo {title} {{On Effective Degrees of Freedom in the Early Universe}},}\ }\href {\doibase 10.3390/galaxies4040078} {\bibfield  {journal} {\bibinfo  {journal} {Galaxies}\ }\textbf {\bibinfo {volume} {4}},\ \bibinfo {pages} {78} (\bibinfo {year} {2016})},\ \Eprint {http://arxiv.org/abs/1609.04979} {arXiv:1609.04979 [astro-ph.CO]} \BibitemShut {NoStop}%
\bibitem [{\citenamefont {Srednicki}\ \emph {et~al.}(1988)\citenamefont {Srednicki}, \citenamefont {Watkins},\ and\ \citenamefont {Olive}}]{Srednicki:1988ce}%
  \BibitemOpen
  \bibfield  {author} {\bibinfo {author} {\bibfnamefont {Mark}\ \bibnamefont {Srednicki}}, \bibinfo {author} {\bibfnamefont {Richard}\ \bibnamefont {Watkins}}, \ and\ \bibinfo {author} {\bibfnamefont {Keith~A.}\ \bibnamefont {Olive}},\ }\bibfield  {title} {\enquote {\bibinfo {title} {{Calculations of Relic Densities in the Early Universe}},}\ }\href {\doibase 10.1016/0550-3213(88)90099-5} {\bibfield  {journal} {\bibinfo  {journal} {Nucl. Phys. B}\ }\textbf {\bibinfo {volume} {310}},\ \bibinfo {pages} {693} (\bibinfo {year} {1988})}\BibitemShut {NoStop}%
\bibitem [{\citenamefont {Kolb}\ and\ \citenamefont {Turner}(2019)}]{Kolb:1990vq}%
  \BibitemOpen
  \bibfield  {author} {\bibinfo {author} {\bibfnamefont {Edward~W.}\ \bibnamefont {Kolb}}\ and\ \bibinfo {author} {\bibfnamefont {Michael~S.}\ \bibnamefont {Turner}},\ }\href {\doibase 10.1201/9780429492860} {\emph {\bibinfo {title} {{The Early Universe}}}},\ Vol.~\bibinfo {volume} {69}\ (\bibinfo  {publisher} {Taylor and Francis},\ \bibinfo {year} {2019})\BibitemShut {NoStop}%
\bibitem [{\citenamefont {Slatyer}\ \emph {et~al.}(2009)\citenamefont {Slatyer}, \citenamefont {Padmanabhan},\ and\ \citenamefont {Finkbeiner}}]{Slatyer:2009yq}%
  \BibitemOpen
  \bibfield  {author} {\bibinfo {author} {\bibfnamefont {Tracy~R.}\ \bibnamefont {Slatyer}}, \bibinfo {author} {\bibfnamefont {Nikhil}\ \bibnamefont {Padmanabhan}}, \ and\ \bibinfo {author} {\bibfnamefont {Douglas~P.}\ \bibnamefont {Finkbeiner}},\ }\bibfield  {title} {\enquote {\bibinfo {title} {{CMB Constraints on WIMP Annihilation: Energy Absorption During the Recombination Epoch}},}\ }\href {\doibase 10.1103/PhysRevD.80.043526} {\bibfield  {journal} {\bibinfo  {journal} {Phys. Rev. D}\ }\textbf {\bibinfo {volume} {80}},\ \bibinfo {pages} {043526} (\bibinfo {year} {2009})},\ \Eprint {http://arxiv.org/abs/0906.1197} {arXiv:0906.1197 [astro-ph.CO]} \BibitemShut {NoStop}%
\bibitem [{\citenamefont {Berlin}\ \emph {et~al.}(2024)\citenamefont {Berlin}, \citenamefont {Krnjaic},\ and\ \citenamefont {Pinetti}}]{Berlin:2023qco}%
  \BibitemOpen
  \bibfield  {author} {\bibinfo {author} {\bibfnamefont {Asher}\ \bibnamefont {Berlin}}, \bibinfo {author} {\bibfnamefont {Gordan}\ \bibnamefont {Krnjaic}}, \ and\ \bibinfo {author} {\bibfnamefont {Elena}\ \bibnamefont {Pinetti}},\ }\bibfield  {title} {\enquote {\bibinfo {title} {{Reviving MeV-GeV indirect detection with inelastic dark matter}},}\ }\href {\doibase 10.1103/PhysRevD.110.035015} {\bibfield  {journal} {\bibinfo  {journal} {Phys. Rev. D}\ }\textbf {\bibinfo {volume} {110}},\ \bibinfo {pages} {035015} (\bibinfo {year} {2024})},\ \Eprint {http://arxiv.org/abs/2311.00032} {arXiv:2311.00032 [hep-ph]} \BibitemShut {NoStop}%
\bibitem [{\citenamefont {Aprile}\ \emph {et~al.}(2019)\citenamefont {Aprile} \emph {et~al.}}]{XENON:2019gfn}%
  \BibitemOpen
  \bibfield  {author} {\bibinfo {author} {\bibfnamefont {E.}~\bibnamefont {Aprile}} \emph {et~al.} (\bibinfo {collaboration} {XENON}),\ }\bibfield  {title} {\enquote {\bibinfo {title} {{Light Dark Matter Search with Ionization Signals in XENON1T}},}\ }\href {\doibase 10.1103/PhysRevLett.123.251801} {\bibfield  {journal} {\bibinfo  {journal} {Phys. Rev. Lett.}\ }\textbf {\bibinfo {volume} {123}},\ \bibinfo {pages} {251801} (\bibinfo {year} {2019})},\ \Eprint {http://arxiv.org/abs/1907.11485} {arXiv:1907.11485 [hep-ex]} \BibitemShut {NoStop}%
\bibitem [{\citenamefont {Li}\ \emph {et~al.}(2023)\citenamefont {Li} \emph {et~al.}}]{PandaX:2022xqx}%
  \BibitemOpen
  \bibfield  {author} {\bibinfo {author} {\bibfnamefont {Shuaijie}\ \bibnamefont {Li}} \emph {et~al.} (\bibinfo {collaboration} {PandaX}),\ }\bibfield  {title} {\enquote {\bibinfo {title} {{Search for Light Dark Matter with Ionization Signals in the PandaX-4T Experiment}},}\ }\href {\doibase 10.1103/PhysRevLett.130.261001} {\bibfield  {journal} {\bibinfo  {journal} {Phys. Rev. Lett.}\ }\textbf {\bibinfo {volume} {130}},\ \bibinfo {pages} {261001} (\bibinfo {year} {2023})},\ \Eprint {http://arxiv.org/abs/2212.10067} {arXiv:2212.10067 [hep-ex]} \BibitemShut {NoStop}%
\bibitem [{\citenamefont {Akerib}\ \emph {et~al.}(2025)\citenamefont {Akerib} \emph {et~al.}}]{LZ:2025zpw}%
  \BibitemOpen
  \bibfield  {author} {\bibinfo {author} {\bibfnamefont {D.~S.}\ \bibnamefont {Akerib}} \emph {et~al.} (\bibinfo {collaboration} {LZ}),\ }\bibfield  {title} {\enquote {\bibinfo {title} {{Search for New Physics via Low-Energy Electron Recoils with a 4.2 Tonne{\texttimes} Year Exposure from the LZ Experiment}},}\ }\href@noop {} {\  (\bibinfo {year} {2025})},\ \Eprint {http://arxiv.org/abs/2511.17350} {arXiv:2511.17350 [hep-ex]} \BibitemShut {NoStop}%
\bibitem [{\citenamefont {Smith}\ \emph {et~al.}(2007)\citenamefont {Smith} \emph {et~al.}}]{Smith:2006ym}%
  \BibitemOpen
  \bibfield  {author} {\bibinfo {author} {\bibfnamefont {Martin~C.}\ \bibnamefont {Smith}} \emph {et~al.},\ }\bibfield  {title} {\enquote {\bibinfo {title} {{The RAVE Survey: Constraining the Local Galactic Escape Speed}},}\ }\href {\doibase 10.1111/j.1365-2966.2007.11964.x} {\bibfield  {journal} {\bibinfo  {journal} {Mon. Not. Roy. Astron. Soc.}\ }\textbf {\bibinfo {volume} {379}},\ \bibinfo {pages} {755--772} (\bibinfo {year} {2007})},\ \Eprint {http://arxiv.org/abs/astro-ph/0611671} {arXiv:astro-ph/0611671} \BibitemShut {NoStop}%
\bibitem [{\citenamefont {Green}(2017)}]{Green:2017odb}%
  \BibitemOpen
  \bibfield  {author} {\bibinfo {author} {\bibfnamefont {Anne~M}\ \bibnamefont {Green}},\ }\bibfield  {title} {\enquote {\bibinfo {title} {{Astrophysical uncertainties on the local dark matter distribution and direct detection experiments}},}\ }\href {\doibase 10.1088/1361-6471/aa7819} {\bibfield  {journal} {\bibinfo  {journal} {J. Phys. G}\ }\textbf {\bibinfo {volume} {44}},\ \bibinfo {pages} {084001} (\bibinfo {year} {2017})},\ \Eprint {http://arxiv.org/abs/1703.10102} {arXiv:1703.10102 [astro-ph.CO]} \BibitemShut {NoStop}%
\bibitem [{\citenamefont {Krnjaic}\ \emph {et~al.}(2025{\natexlab{b}})\citenamefont {Krnjaic}, \citenamefont {Rocha},\ and\ \citenamefont {Trickle}}]{Krnjaic:2024bdd}%
  \BibitemOpen
  \bibfield  {author} {\bibinfo {author} {\bibfnamefont {Gordan}\ \bibnamefont {Krnjaic}}, \bibinfo {author} {\bibfnamefont {Duncan}\ \bibnamefont {Rocha}}, \ and\ \bibinfo {author} {\bibfnamefont {Tanner}\ \bibnamefont {Trickle}},\ }\bibfield  {title} {\enquote {\bibinfo {title} {{The non-relativistic effective field theory of dark matter-electron interactions}},}\ }\href {\doibase 10.1007/JHEP03(2025)165} {\bibfield  {journal} {\bibinfo  {journal} {JHEP}\ }\textbf {\bibinfo {volume} {03}},\ \bibinfo {pages} {165} (\bibinfo {year} {2025}{\natexlab{b}})},\ \Eprint {http://arxiv.org/abs/2407.14598} {arXiv:2407.14598 [hep-ph]} \BibitemShut {NoStop}%
\bibitem [{\citenamefont {Catena}\ \emph {et~al.}(2020)\citenamefont {Catena}, \citenamefont {Emken}, \citenamefont {Spaldin},\ and\ \citenamefont {Tarantino}}]{Catena:2019gfa}%
  \BibitemOpen
  \bibfield  {author} {\bibinfo {author} {\bibfnamefont {Riccardo}\ \bibnamefont {Catena}}, \bibinfo {author} {\bibfnamefont {Timon}\ \bibnamefont {Emken}}, \bibinfo {author} {\bibfnamefont {Nicola~A.}\ \bibnamefont {Spaldin}}, \ and\ \bibinfo {author} {\bibfnamefont {Walter}\ \bibnamefont {Tarantino}},\ }\bibfield  {title} {\enquote {\bibinfo {title} {{Atomic responses to general dark matter-electron interactions}},}\ }\href {\doibase 10.1103/PhysRevResearch.2.033195} {\bibfield  {journal} {\bibinfo  {journal} {Phys. Rev. Res.}\ }\textbf {\bibinfo {volume} {2}},\ \bibinfo {pages} {033195} (\bibinfo {year} {2020})},\ \bibinfo {note} {[Erratum: Phys.Rev.Res. 7, 019001 (2025)]},\ \Eprint {http://arxiv.org/abs/1912.08204} {arXiv:1912.08204 [hep-ph]} \BibitemShut {NoStop}%
\bibitem [{\citenamefont {Liang}\ \emph {et~al.}(2024)\citenamefont {Liang}, \citenamefont {Liao}, \citenamefont {Ma},\ and\ \citenamefont {Wang}}]{Liang:2024ecw}%
  \BibitemOpen
  \bibfield  {author} {\bibinfo {author} {\bibfnamefont {Jin-Han}\ \bibnamefont {Liang}}, \bibinfo {author} {\bibfnamefont {Yi}~\bibnamefont {Liao}}, \bibinfo {author} {\bibfnamefont {Xiao-Dong}\ \bibnamefont {Ma}}, \ and\ \bibinfo {author} {\bibfnamefont {Hao-Lin}\ \bibnamefont {Wang}},\ }\bibfield  {title} {\enquote {\bibinfo {title} {{A systematic investigation on dark matter-electron scattering in effective field theories}},}\ }\href {\doibase 10.1007/JHEP07(2024)279} {\bibfield  {journal} {\bibinfo  {journal} {JHEP}\ }\textbf {\bibinfo {volume} {07}},\ \bibinfo {pages} {279} (\bibinfo {year} {2024})},\ \Eprint {http://arxiv.org/abs/2406.10912} {arXiv:2406.10912 [hep-ph]} \BibitemShut {NoStop}%
\bibitem [{\citenamefont {Read}(2014)}]{Read:2014qva}%
  \BibitemOpen
  \bibfield  {author} {\bibinfo {author} {\bibfnamefont {J.~I.}\ \bibnamefont {Read}},\ }\bibfield  {title} {\enquote {\bibinfo {title} {{The Local Dark Matter Density}},}\ }\href {\doibase 10.1088/0954-3899/41/6/063101} {\bibfield  {journal} {\bibinfo  {journal} {J. Phys. G}\ }\textbf {\bibinfo {volume} {41}},\ \bibinfo {pages} {063101} (\bibinfo {year} {2014})},\ \Eprint {http://arxiv.org/abs/1404.1938} {arXiv:1404.1938 [astro-ph.GA]} \BibitemShut {NoStop}%
\bibitem [{\citenamefont {Savage}\ \emph {et~al.}(2006)\citenamefont {Savage}, \citenamefont {Freese},\ and\ \citenamefont {Gondolo}}]{Savage:2006qr}%
  \BibitemOpen
  \bibfield  {author} {\bibinfo {author} {\bibfnamefont {Christopher}\ \bibnamefont {Savage}}, \bibinfo {author} {\bibfnamefont {Katherine}\ \bibnamefont {Freese}}, \ and\ \bibinfo {author} {\bibfnamefont {Paolo}\ \bibnamefont {Gondolo}},\ }\bibfield  {title} {\enquote {\bibinfo {title} {{Annual Modulation of Dark Matter in the Presence of Streams}},}\ }\href {\doibase 10.1103/PhysRevD.74.043531} {\bibfield  {journal} {\bibinfo  {journal} {Phys. Rev. D}\ }\textbf {\bibinfo {volume} {74}},\ \bibinfo {pages} {043531} (\bibinfo {year} {2006})},\ \Eprint {http://arxiv.org/abs/astro-ph/0607121} {arXiv:astro-ph/0607121} \BibitemShut {NoStop}%
\bibitem [{\citenamefont {McCabe}(2010)}]{McCabe:2010zh}%
  \BibitemOpen
  \bibfield  {author} {\bibinfo {author} {\bibfnamefont {Christopher}\ \bibnamefont {McCabe}},\ }\bibfield  {title} {\enquote {\bibinfo {title} {{The Astrophysical Uncertainties Of Dark Matter Direct Detection Experiments}},}\ }\href {\doibase 10.1103/PhysRevD.82.023530} {\bibfield  {journal} {\bibinfo  {journal} {Phys. Rev. D}\ }\textbf {\bibinfo {volume} {82}},\ \bibinfo {pages} {023530} (\bibinfo {year} {2010})},\ \Eprint {http://arxiv.org/abs/1005.0579} {arXiv:1005.0579 [hep-ph]} \BibitemShut {NoStop}%
\bibitem [{\citenamefont {Yi}\ \emph {et~al.}(2026)\citenamefont {Yi} \emph {et~al.}}]{Yi:2026fmf}%
  \BibitemOpen
  \bibfield  {author} {\bibinfo {author} {\bibfnamefont {Difan}\ \bibnamefont {Yi}} \emph {et~al.},\ }\bibfield  {title} {\enquote {\bibinfo {title} {{Direct observation of the Migdal effect induced by neutron bombardment}},}\ }\href {\doibase 10.1038/s41586-025-09918-8} {\bibfield  {journal} {\bibinfo  {journal} {Nature}\ }\textbf {\bibinfo {volume} {649}},\ \bibinfo {pages} {580--583} (\bibinfo {year} {2026})}\BibitemShut {NoStop}%
\bibitem [{\citenamefont {Ibe}\ \emph {et~al.}(2018)\citenamefont {Ibe}, \citenamefont {Nakano}, \citenamefont {Shoji},\ and\ \citenamefont {Suzuki}}]{Ibe:2017yqa}%
  \BibitemOpen
  \bibfield  {author} {\bibinfo {author} {\bibfnamefont {Masahiro}\ \bibnamefont {Ibe}}, \bibinfo {author} {\bibfnamefont {Wakutaka}\ \bibnamefont {Nakano}}, \bibinfo {author} {\bibfnamefont {Yutaro}\ \bibnamefont {Shoji}}, \ and\ \bibinfo {author} {\bibfnamefont {Kazumine}\ \bibnamefont {Suzuki}},\ }\bibfield  {title} {\enquote {\bibinfo {title} {{Migdal Effect in Dark Matter Direct Detection Experiments}},}\ }\href {\doibase 10.1007/JHEP03(2018)194} {\bibfield  {journal} {\bibinfo  {journal} {JHEP}\ }\textbf {\bibinfo {volume} {03}},\ \bibinfo {pages} {194} (\bibinfo {year} {2018})},\ \Eprint {http://arxiv.org/abs/1707.07258} {arXiv:1707.07258 [hep-ph]} \BibitemShut {NoStop}%
\bibitem [{\citenamefont {Baxter}\ \emph {et~al.}(2021)\citenamefont {Baxter} \emph {et~al.}}]{Baxter:2021pqo}%
  \BibitemOpen
  \bibfield  {author} {\bibinfo {author} {\bibfnamefont {D.}~\bibnamefont {Baxter}} \emph {et~al.},\ }\bibfield  {title} {\enquote {\bibinfo {title} {{Recommended conventions for reporting results from direct dark matter searches}},}\ }\href {\doibase 10.1140/epjc/s10052-021-09655-y} {\bibfield  {journal} {\bibinfo  {journal} {Eur. Phys. J. C}\ }\textbf {\bibinfo {volume} {81}},\ \bibinfo {pages} {907} (\bibinfo {year} {2021})},\ \Eprint {http://arxiv.org/abs/2105.00599} {arXiv:2105.00599 [hep-ex]} \BibitemShut {NoStop}%
\bibitem [{\citenamefont {Magill}\ \emph {et~al.}(2018)\citenamefont {Magill}, \citenamefont {Plestid}, \citenamefont {Pospelov},\ and\ \citenamefont {Tsai}}]{Magill:2018jla}%
  \BibitemOpen
  \bibfield  {author} {\bibinfo {author} {\bibfnamefont {Gabriel}\ \bibnamefont {Magill}}, \bibinfo {author} {\bibfnamefont {Ryan}\ \bibnamefont {Plestid}}, \bibinfo {author} {\bibfnamefont {Maxim}\ \bibnamefont {Pospelov}}, \ and\ \bibinfo {author} {\bibfnamefont {Yu-Dai}\ \bibnamefont {Tsai}},\ }\bibfield  {title} {\enquote {\bibinfo {title} {{Dipole Portal to Heavy Neutral Leptons}},}\ }\href {\doibase 10.1103/PhysRevD.98.115015} {\bibfield  {journal} {\bibinfo  {journal} {Phys. Rev. D}\ }\textbf {\bibinfo {volume} {98}},\ \bibinfo {pages} {115015} (\bibinfo {year} {2018})},\ \Eprint {http://arxiv.org/abs/1803.03262} {arXiv:1803.03262 [hep-ph]} \BibitemShut {NoStop}%
\bibitem [{\citenamefont {Magill}\ \emph {et~al.}(2019)\citenamefont {Magill}, \citenamefont {Plestid}, \citenamefont {Pospelov},\ and\ \citenamefont {Tsai}}]{Magill:2018tbb}%
  \BibitemOpen
  \bibfield  {author} {\bibinfo {author} {\bibfnamefont {Gabriel}\ \bibnamefont {Magill}}, \bibinfo {author} {\bibfnamefont {Ryan}\ \bibnamefont {Plestid}}, \bibinfo {author} {\bibfnamefont {Maxim}\ \bibnamefont {Pospelov}}, \ and\ \bibinfo {author} {\bibfnamefont {Yu-Dai}\ \bibnamefont {Tsai}},\ }\bibfield  {title} {\enquote {\bibinfo {title} {{Millicharged particles in neutrino experiments}},}\ }\href {\doibase 10.1103/PhysRevLett.122.071801} {\bibfield  {journal} {\bibinfo  {journal} {Phys. Rev. Lett.}\ }\textbf {\bibinfo {volume} {122}},\ \bibinfo {pages} {071801} (\bibinfo {year} {2019})},\ \Eprint {http://arxiv.org/abs/1806.03310} {arXiv:1806.03310 [hep-ph]} \BibitemShut {NoStop}%
\bibitem [{\citenamefont {Navas}\ \emph {et~al.}(2024)\citenamefont {Navas} \emph {et~al.}}]{ParticleDataGroup:2024cfk}%
  \BibitemOpen
  \bibfield  {author} {\bibinfo {author} {\bibfnamefont {S.}~\bibnamefont {Navas}} \emph {et~al.} (\bibinfo {collaboration} {Particle Data Group}),\ }\bibfield  {title} {\enquote {\bibinfo {title} {{Review of particle physics}},}\ }\href {\doibase 10.1103/PhysRevD.110.030001} {\bibfield  {journal} {\bibinfo  {journal} {Phys. Rev. D}\ }\textbf {\bibinfo {volume} {110}},\ \bibinfo {pages} {030001} (\bibinfo {year} {2024})}\BibitemShut {NoStop}%
\bibitem [{\citenamefont {Cowan}\ \emph {et~al.}(2011)\citenamefont {Cowan}, \citenamefont {Cranmer}, \citenamefont {Gross},\ and\ \citenamefont {Vitells}}]{Cowan:2010js}%
  \BibitemOpen
  \bibfield  {author} {\bibinfo {author} {\bibfnamefont {Glen}\ \bibnamefont {Cowan}}, \bibinfo {author} {\bibfnamefont {Kyle}\ \bibnamefont {Cranmer}}, \bibinfo {author} {\bibfnamefont {Eilam}\ \bibnamefont {Gross}}, \ and\ \bibinfo {author} {\bibfnamefont {Ofer}\ \bibnamefont {Vitells}},\ }\bibfield  {title} {\enquote {\bibinfo {title} {{Asymptotic formulae for likelihood-based tests of new physics}},}\ }\href {\doibase 10.1140/epjc/s10052-011-1554-0} {\bibfield  {journal} {\bibinfo  {journal} {Eur. Phys. J. C}\ }\textbf {\bibinfo {volume} {71}},\ \bibinfo {pages} {1554} (\bibinfo {year} {2011})},\ \bibinfo {note} {[Erratum: Eur.Phys.J.C 73, 2501 (2013)]},\ \Eprint {http://arxiv.org/abs/1007.1727} {arXiv:1007.1727 [physics.data-an]} \BibitemShut {NoStop}%
\bibitem [{\citenamefont {Agnes}\ \emph {et~al.}(2023)\citenamefont {Agnes} \emph {et~al.}}]{DarkSide-50:2023fcw}%
  \BibitemOpen
  \bibfield  {author} {\bibinfo {author} {\bibfnamefont {P.}~\bibnamefont {Agnes}} \emph {et~al.} (\bibinfo {collaboration} {DarkSide-50}),\ }\bibfield  {title} {\enquote {\bibinfo {title} {{Search for low mass dark matter in DarkSide-50: the bayesian network approach}},}\ }\href {\doibase 10.1140/epjc/s10052-023-11410-4} {\bibfield  {journal} {\bibinfo  {journal} {Eur. Phys. J. C}\ }\textbf {\bibinfo {volume} {83}},\ \bibinfo {pages} {322} (\bibinfo {year} {2023})},\ \Eprint {http://arxiv.org/abs/2302.01830} {arXiv:2302.01830 [hep-ex]} \BibitemShut {NoStop}%
\bibitem [{\citenamefont {Lees}\ \emph {et~al.}(2017)\citenamefont {Lees} \emph {et~al.}}]{BaBar:2017tiz}%
  \BibitemOpen
  \bibfield  {author} {\bibinfo {author} {\bibfnamefont {J.~P.}\ \bibnamefont {Lees}} \emph {et~al.} (\bibinfo {collaboration} {BaBar}),\ }\bibfield  {title} {\enquote {\bibinfo {title} {{Search for Invisible Decays of a Dark Photon Produced in ${e}^{+}{e}^{-}$ Collisions at BaBar}},}\ }\href {\doibase 10.1103/PhysRevLett.119.131804} {\bibfield  {journal} {\bibinfo  {journal} {Phys. Rev. Lett.}\ }\textbf {\bibinfo {volume} {119}},\ \bibinfo {pages} {131804} (\bibinfo {year} {2017})},\ \Eprint {http://arxiv.org/abs/1702.03327} {arXiv:1702.03327 [hep-ex]} \BibitemShut {NoStop}%
\bibitem [{\citenamefont {Krnjaic}\ \emph {et~al.}(2020)\citenamefont {Krnjaic}, \citenamefont {Marques-Tavares}, \citenamefont {Redigolo},\ and\ \citenamefont {Tobioka}}]{Krnjaic:2019rsv}%
  \BibitemOpen
  \bibfield  {author} {\bibinfo {author} {\bibfnamefont {Gordan}\ \bibnamefont {Krnjaic}}, \bibinfo {author} {\bibfnamefont {Gustavo}\ \bibnamefont {Marques-Tavares}}, \bibinfo {author} {\bibfnamefont {Diego}\ \bibnamefont {Redigolo}}, \ and\ \bibinfo {author} {\bibfnamefont {Kohsaku}\ \bibnamefont {Tobioka}},\ }\bibfield  {title} {\enquote {\bibinfo {title} {{Probing Muonphilic Force Carriers and Dark Matter at Kaon Factories}},}\ }\href {\doibase 10.1103/PhysRevLett.124.041802} {\bibfield  {journal} {\bibinfo  {journal} {Phys. Rev. Lett.}\ }\textbf {\bibinfo {volume} {124}},\ \bibinfo {pages} {041802} (\bibinfo {year} {2020})},\ \Eprint {http://arxiv.org/abs/1902.07715} {arXiv:1902.07715 [hep-ph]} \BibitemShut {NoStop}%
\bibitem [{\citenamefont {Cortina~Gil}\ \emph {et~al.}(2021)\citenamefont {Cortina~Gil} \emph {et~al.}}]{NA62:2021bji}%
  \BibitemOpen
  \bibfield  {author} {\bibinfo {author} {\bibfnamefont {Eduardo}\ \bibnamefont {Cortina~Gil}} \emph {et~al.} (\bibinfo {collaboration} {NA62}),\ }\bibfield  {title} {\enquote {\bibinfo {title} {{Search for $K^+$ decays to a muon and invisible particles}},}\ }\href {\doibase 10.1016/j.physletb.2021.136259} {\bibfield  {journal} {\bibinfo  {journal} {Phys. Lett. B}\ }\textbf {\bibinfo {volume} {816}},\ \bibinfo {pages} {136259} (\bibinfo {year} {2021})},\ \Eprint {http://arxiv.org/abs/2101.12304} {arXiv:2101.12304 [hep-ex]} \BibitemShut {NoStop}%
\end{thebibliography}%

\end{document}